\documentclass[12pt, draftclsnofoot, onecolumn]{IEEEtran}

\usepackage[english]{babel}

\usepackage[letterpaper,top=2cm,bottom=2cm,left=3cm,right=3cm,marginparwidth=1.75cm]{geometry}

\usepackage{amsmath,amsfonts}
\usepackage{graphicx}
\usepackage[colorlinks=true, allcolors=blue]{hyperref}
\usepackage{bbm}
\usepackage{caption,subcaption}
\usepackage{verbatim}
\usepackage{cite,enumerate}
\usepackage{srcltx,cases}
\usepackage[mathscr]{eucal}

\newtheorem{theorem}{Theorem}
\newtheorem{lemma}{Lemma}

\newtheorem{corollary}{Corollary}

\title{Coverage Analysis of LEO Satellite Downlink Networks:\\ Orbit Geometry Dependent Approach}
\author{Junse Lee, Song Noh, Sooyeob Jeong, Namyoon Lee\thanks{Junse Lee is with School of AI Convergence, Sungshin Women's University, 02844 Seoul, South Korea (e-mail: junselee@sungshin.ac.kr).}\thanks{Song Noh is with Department of Information and Telecommunication Engineering, Incheon National University, 22012 Incheon, South Korea (e-mail: songnoh@inu.ac.kr).}\thanks{Sooyeob Jeong is with Satellite Wide-Area Infra Research Section, Electronics and Telecommunications Research Institute, ETRI, 34129 Daejeon, South Korea (e-mail: jung2816@etri.re.kr).}\thanks{Namyoon Lee is with School of Electrical Engineering, Korea University, 02841 Seoul, South Korea (e-mail: namyoon@korea.ac.kr).}\thanks{J. Lee and S. Jeong were supported by Institute of Information $\&$ communications Technology Planning $\&$ Evaluation (IITP) grant funded by the Korea government (MSIT) (No.2020-0-00843, Development of low power satellite multiple access core technology based on LEO cubesat for global IoT service).  N. Lee is supported by the National Research Foundation of Korea (NRF) grant funded by the Korea government (MSIT) (No. 2020R1C1C1013381).}}

\begin{document}
\maketitle

\begin{abstract}
The low-earth-orbit (LEO) satellite network with mega-constellations can provide global coverage while supporting the high-data rates. The coverage performance of such a network is highly dependent on orbit geometry parameters, including satellite altitude and inclination angle. Traditionally, simulation-based coverage analysis dominates because of the lack of analytical approaches. This paper presents a novel systematic analysis framework for the LEO satellite network by highlighting orbit geometric parameters. Specifically, we assume that satellite locations are placed on a circular orbit according to a one-dimensional Poisson point process. Then, we derive the distribution of the nearest distance between the satellite and a fixed user's location on the Earth in terms of the orbit-geometry parameters. Leveraging this distribution, we characterize the coverage probability of the single-orbit LEO network as a function of the network geometric parameters in conjunction with small and large-scale fading effects. Finally, we extend our coverage analysis to multi-orbit networks and verify the synergistic gain of harnessing multi-orbit satellite networks in terms of the coverage probability. Simulation results are provided to validate the mathematical derivations and the accuracy of the proposed model.

\end{abstract}


\IEEEpeerreviewmaketitle
\section{Introduction}
\subsection{Motivation}
The appetite for seamless global coverage is unappeasable for the next generation of communication systems \cite{chen2020vision,zhang20196g,giordani2020toward}. Despite the broad coverage areas of terrestrial cellular networks, including from 1G to 5G, half of the global population are still under-connected to the Internet \cite{international2019measuring}. In addition, aero and maritime lanes remain unconnected because they are non-covered areas by terrestrial networks. 

Satellite networks can provide a ubiquitous footprint across the globe. For instance, it is possible to provide seamless coverage across the Earth with only three geostationary Earth orbit (GEO) satellites. Yet, these GEO satellite networks are very limited to use for 5G applications due to extremely low spectral efficiency and tremendously high latency. An emerging alternative is LEO satellite networks with massive constellations.

The LEO satellites can be densely deployed with mega-constellations thanks to cost-effective launch systems, providing considerably higher data rates and lower latency than their classical GEO counterparts. For instance, Starlink is targeting to deploy around 42000 LEO satellites on several different altitudes of orbits between 330 km and  570 km. OneWeb also has deployed around 100 LEO satellites to provide seamless global coverage. This LEO satellite network with mage-constellation can support high data rates and low latency to meet the stringent requirements for 5G applications while providing a global footprint. 

Understanding the coverage of LEO satellite networks is significant in terms of the relevant system parameters to optimize network planning, i.e., where to place new satellites to maximize coverage while minimizing cost. Traditionally, complex system-level simulations have been used to optimize satellite deployment. For instance, the Walker constellation, placing the satellites on a gird of multiple orbits, has been commonly used to evaluate the coverage performance of LEO satellite networks. However, this simulation-based study is limited to understanding the interplay among many network design parameters. Hence, analytical modeling and analysis tool is essential to provide network design insights as an alternative approach.

Stochastic geometry is a mathematical tool that analyzes the spatially averaged performance of wireless networks. By modeling the locations of transmitters and receivers according to proper point processes, stochastic geometry can provide insights into the relationship between the network performance and parameters. For instance, this tool has made a successful progress on characterizing the coverage and rate performance for ad-hoc models\cite{baccelli2009stochastic,baccelli2006aloha,baccelli2009stochasticopp,NLee3}, cellular models\cite{andrews2011tractable,dhillon2012modeling,di2013average,NLee1,NLee2}, mmWave\cite{bai2014coverage,di2015stochastic}, V2X\cite{tong2016stochastic,yi2019modeling}, and UAV models\cite{chetlur2017downlink,banagar2020performance}. Recently, it has been applied to analyze LEO satellite networks with mega-constellations \cite{okati2021modeling,okati2022coverage,okati2022nonhomogeneous,al2022next}. 

Continuing the success of the prior work, in this paper, we develop a novel analytical framework for the coverage analysis of downlink satellite networks. Unlike the previous approach in \cite{okati2021modeling,okati2022coverage,okati2022nonhomogeneous,talgat2020nearest,talgat2020stochastic,al2021analytic,al2021optimal,park2021coverage}, we characterize the downlink coverage probability of the LEO satellite networks by highlighting the satellite orbit geometry relative to the receiver's orientation.



\subsection{Related works}
Modeling satellite networks using stochastic geometry has gained momentum because of its tractability in the coverage analysis compared to using the classical grid models such as Walker's constellation \cite{walker1970circular,ganz1994performance,vatalaro1995analysis}. The location of the satellite is modeled by placing a homogeneous binomial point process (BPP) on spherical surfaces in \cite{okati2020downlink,talgat2020nearest,talgat2020stochastic}. Under the premise that a fixed number of satellites are uniformly distributed on the surface of a sphere, the prior work in  \cite{okati2020downlink} established the coverage probability. Also, in \cite{talgat2020nearest}, the authors characterized the distribution of the distance of two links: 1) between the user and the nearest satellite and 2) between the satellite and the nearest satellite when satellites are assumed to be located at different altitudes. Based on the results in \cite{talgat2020nearest},  the downlink coverage probability was derived in \cite{talgat2020stochastic} when the satellite operates as a relay between the user and other satellites.


Modeling a spatial distribution of satellites according to PPPs is also popular approach \cite{al2021analytic,al2021optimal,okati2021modeling}. Unlike BPP, the number of satellites is assumed to follow a Poisson distribution. In \cite{al2021analytic} and \cite{al2021optimal}, the coverage probability in the satellite network was derived using the approximated contact angle distribution, and the optimal network parameters were attained. \cite{okati2021modeling} and \cite{okati2022nonhomogeneous} analyze the coverage under modeling the satellite network as a non-homogeneous PPP because the density of satellites viewed by users differs according to latitude. Recently, a tractable approach for the coverage analysis was proposed using a homogeneous PPP in \cite{park2021coverage}. Unlike the literature in \cite{al2021analytic,al2021optimal,okati2021modeling}, in \cite{park2021coverage} a two-step computation method was presented for the coverage analysis; it first computes the coverage probability conditioned on that at least one satellite exists in the field of view of a downlink user and then marginalizes it with respect to the satellite visible distribution. Using this two-step method, the tractable expressions for the coverage probability with quasi-closed forms were derived as a function of the relevant network parameters.

The previous coverage analysis results using BPPs and PPPs in a spherical cap can provide useful guidance in understanding the spatially averaged coverage performance for a densely deployed LEO satellite network. However, they lack understanding of the coverage performance for a specific orbit geometry relative to the receiver's location. In this paper, we take a different approach from the existing spatial modeling of satellite locations. We model that satellite locations follow a homogeneous PPP on a fixed circular orbit, i.e., a one-dimensional PPP on a line. This spatial modeling can better capture the physical movements of the satellites than the spatial distributions using two-dimensional BPPs and PPPs. In addition, our satellite spatial modeling allows us to understand the coverage performance in terms of the orbit geometry parameters relative to the receiver's location. Therefore, our analysis is helpful in designing orbit parameters, which disappear when taking the previous methods in \cite{al2021analytic,al2021optimal,okati2021modeling, park2021coverage}.

\subsection{Contributions}
This paper puts forth a novel coverage analysis framework for satellite downlink networks consisting of a given set of satellite orbits. Capitalizing on the given circular orbit parameters relative to a fixed receiver's location, we model that the satellite locations as independent homogeneous PPPs on each orbit. The main contributions of this paper are summarized as follows:
\begin{itemize}
    \item We first consider the distribution of visible orbits and satellites by the user. First, by mapping an orbit to a length of one vector, the length of an orbit on a visible surface is characterized. Then, we obtain the distribution of visible satellites by the user. Unlike the previous approaches in \cite{okati2021modeling,okati2022coverage,talgat2020nearest,talgat2020stochastic,al2021analytic,al2021optimal,park2021coverage}, the satellites are distributed according to a union of one-dimensional PPPs in a finite area.  From the derived satellite visible probability, we confirm that the satellite is more visible when its relative angle from the receiver is low, confirming our intuition that there is the highest chance to see the satellites in orbit with a zero angle from the receiver. 

    \item  Leveraging the derived satellite visibility probability, we derive a distribution of the distance between the user and its nearest satellite as a function of network parameters conditioned on the existence of at least one satellite visible by the user in each orbit. One remarkable observation is that this nearest distance distribution differs from the counterpart when modeling the satellite locations according to homogeneous PPPs or BPPs in the spherical cap, i.e., the two-dimensional space. The complementary cumulative distribution function (CCDF) shows that the nearest distance distribution improves as the network density increases while the relative orbit angle decreases.

    \item Using the contact distance distribution, which is an important segment in deriving the coverage probability of a satellite downlink network, we characterize an analytical expression for the coverage probability for a single-orbit LEO satellite network in terms of the satellite orbit angle and fading parameters. From our coverage analysis, we confirm that the coverage probability is inversely proportional to both the satellite density, orbit angle, and orbit altitudes, provided that at least one satellite is visible. 
    

    \item We extend our coverage probability expression when a receiver user can communicate with satellites on multiple orbits, each with orthogonal time-frequency resources. Using the maximum signal-to-interference ratio (SIR) selection strategy, in which the receiver selects the satellite on the orbit, which provides the maximum SIR, we establish the coverage probability expression in terms of the number of orbits. We verify that the coverage probability significantly improves as the number of orbits increases, and the coverage probability enhancement by harnessing more orbits is proportional to the length of the satellite's visible trajectory. From simulations, we verify the accuracy of the derived coverage probability expressions. 
\end{itemize}




The remainder of this paper is organized as follows. After presenting the system model and coverage probability in Section \ref{sec:Section 2}, we characterize the visible probability of satellites at the fixed position in \ref{sec: Section3}. Then, we derive the coverage probability experienced by the user under the single-orbit network in Section \ref{sec: Section4} and investigate the benefit of the opportunistic diversity scheme under the multi-orbit network in Section \ref{sec:multiple_orbits} before concluding this paper in Section \ref{sec: Section6}.


\section{System Model}\label{sec:Section 2}
This section explains the network and channel models for LEO downlink satellite networks. Then, we introduce the coverage probability in terms of relevant network and channel model parameters.

\subsection{Network Model}
\subsubsection{Geometries of satellite orbits}
\begin{figure}[t]
\begin{center}
\includegraphics[scale=0.45]{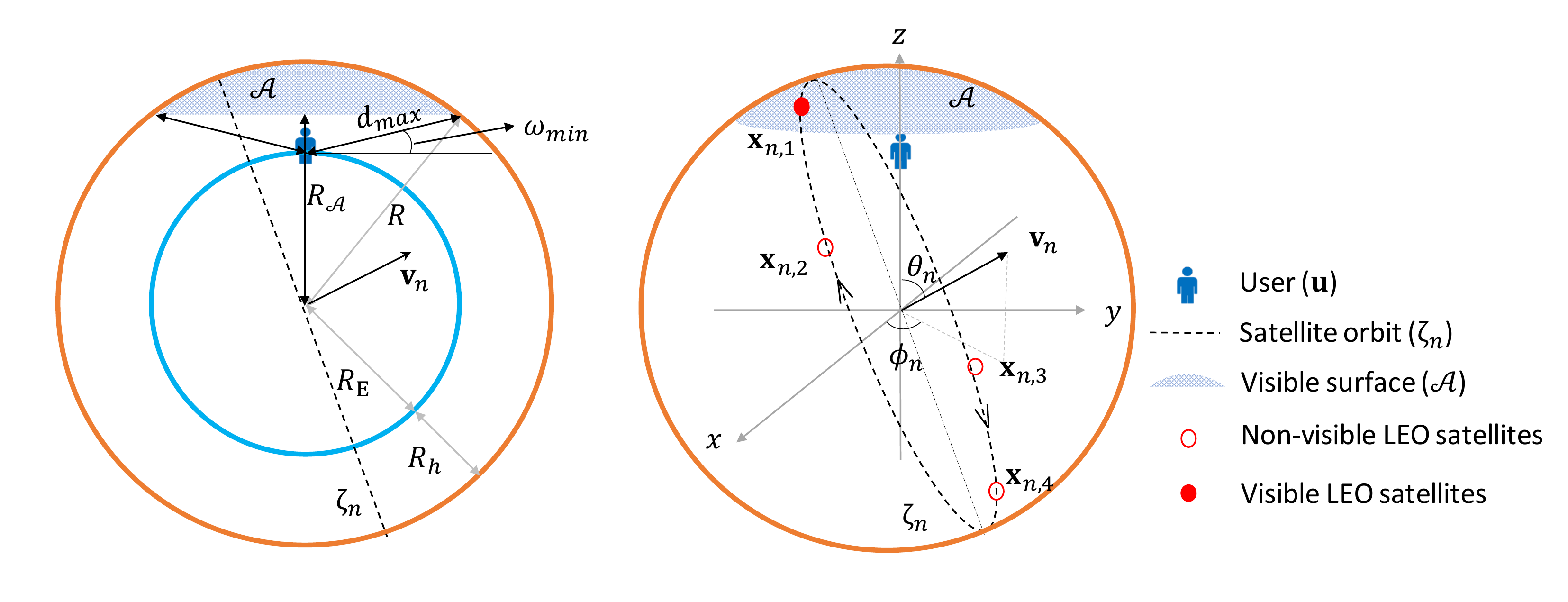}
\end{center}
\vspace{-0.5em} 
\caption{Illustrations of the $n$-th satellite orbit geometry denoted by $\zeta_n$. The dotted line in the left figure depicts an orthogonal projection view of the orbit $\zeta_n$, and the right one illustrates a three-dimensional view of the satellite orbit $\zeta_n$ in terms of relevant parameters, polar angle $\theta_n$ and azimuth angle $\phi_n$ where $M_n=4$. } 
\label{fig:sys} 
\end{figure}

We consider a satellite network comprised of $N$ orbits with the same altitude $R_{\sf h}$. As illustrated in Figure \ref{fig:sys}, we assume that the Earth and all satellite orbits are concentric sphere and circles with radius $R_{\sf E}$ and $R(=R_{\sf E}+R_{\sf h})$. The orbit geometry is completely determined by a normal unit-vector with respect to the plane containing that orbit. Armed with the normal vector $\mathbf{v}_n$, we define the $n$-th orbit geometry with two parameters, polar angle $\theta_n\in [0,\pi]$ and azimuth angle $\phi_n\in [0,2\pi)$ in the spherical coordinate system, where $n\in [N]$.\footnote{$[N]$ denotes the set of integers from 1 to $N$, i.e., $\{1,2,\ldots,N\}$. }

  We denote the $n$-th orbit geometry, i.e., circle, by $\zeta_n$ which is associated with a normal unit-vector $\mathbf{v}_n =(1,\theta_n,\phi_n)$\footnote{In the Cartesian coordinate system, this normal vector is mapped to $(\sin{\theta_n}\cos{\phi_n}, \sin{\theta_n}\sin{\phi_n},\cos{\theta_n})$. } in the spherical coordinate system. For example, satellites on $\zeta_n$ with $\theta_n=\frac{\pi}{2}$ pass $(0,0,R)$ and $(0,0,-R)$, and those with $\theta_n = 0$ orbit on $XY$-plane in the Cartesian coordinate system. In this paper, the Earth's rotation axis and the $z$-axis of the Cartesian coordinate system are not necessarily aligned, but $\theta_n$ can be considered as the inclination angle of $\zeta_n$, provided that these axes are equivalent.

\subsubsection{Spatial distribution of satellites and a user}

Unlike the typical user analysis used in \cite{park2021coverage}, we take a coverage analysis dependent on the user location. For the ease of exposition, we assume that a user is placed at $\mathbf{u}=(0,0, R_{\sf E})\in\mathbb{R}^3$ in the Cartesian coordinate system. We can harness our analysis framework for a different user's location by redefining the orbit geometries relative to the particular user location. Therefore, our analysis framework is universal in the sense of the user's location.

The satellites are assumed to be distributed according to a homogeneous PPP with an intensity $\lambda$ on each orbit. We denote the set of the location of LEO satellites on $\zeta_n$ by $\Phi_{n}=\{\mathbf{x}_{n,1},\ldots,\mathbf{x}_{n,M_n}\}$ where $M_n$ follows a Poisson distribution with mean $2\pi R\lambda$. Let us assume that the indices are sorted based on distance from the user.

Assume that a user can observe satellites that are located above the minimum elevation angle $\omega_{\rm min}$. Then, the visible satellites are placed on a spherical surface as depicted in Figure \ref{fig:sys}. Here, we denote this visible surface at $\mathbf{u}$ by $\mathcal{A}$. The region $\mathcal{A}$ in the Cartesian coordinate is given by
\begin{align}\label{eq:visible_surface_A}
    \mathcal{A}=\{ (x,y,z)\in\mathbb{R}^3 : \{x^2+y^2+z^2 = R^2\} \cap \{z>R_{\mathcal{A}}\}   \}\mbox{,}
\end{align}
where $R_{\mathcal{A}}=d_{\rm max}\sin \omega_{\rm min} +R_{\sf E}$ and
\begin{equation}
    d_{\rm max} = -R_{\sf E} \sin{\omega_{\rm min}} + \sqrt{(R_{\sf E} \sin{\omega_{\rm min}})^2 + 2R_{\sf E} R_{\sf h} + R_{\sf h}^2}\mbox{,}
\end{equation}
which is the maximum distance between $\mathbf{u}$ and $\mathcal{A}$. $R_{\mathcal{A}}$ is the distance between the center of the Earth and the base of $\mathcal{A}$. For the sake of notation, we define the $n$-th arc, i.e., the intersection of the $n$-th orbit and the spherical cap, denoted by ${\bar \zeta}_n=\zeta_n\cap\mathcal{A}$ for $n\in[N]$. In other words, $\bar{\zeta}_n$ is the visible region of $\zeta_n$ by the user located at $\mathbf{u}$. Table I summarizes the geometric parameters used in our satellite network model.

{ \footnotesize
\begin{table}
\begin{center}
\begin{tabular}{|c ||c|} 
 \hline
 \textbf{Symbol} & \textbf{Definition}  \\ 
 \hline
 $R_{\sf E}$, $R_{\sf h}$, $R$  & Radius of the Earth, altitude of satellites, and radius of satellite orbits  \\ 
 \hline
 $\mathbf{u}$ & Location of the user to measure performances: $(0,0,R_{\sf E})$  \\
 \hline
  $\omega_{\rm min}$ & Minimum elevation angle  \\ 
  \hline
 $\mathcal{A}$ & Visible surface on a sphere with a radius $R$ at $\mathbf{u}$  \\  \hline
 $R_{\mathcal{A}}$ & Distance between the center of the Earth and the base of $\mathcal{A}$  \\ 
 \hline
  $d_{\rm max}$ & Maximum distance between $\mathbf{u}$ and $\mathcal{A}$  \\ 
 \hline
 $\zeta_n$, $\mathbf{v}_n$ & The $n$-th orbit and its associated unit normal vector  \\
 \hline
 $\bar{\zeta}_n$ & Visible region of $\zeta_n$ by the user at $\mathbf{u}$  \\
 \hline
 $\theta_n$, $\phi_n$ & Polar angle and azimuth angle of $\mathbf{v}_n$  \\
  \hline
 $\Phi_n$ & All satellites on $\zeta_n$  \\ 
  \hline
 $\lambda$ & Density of satellites on each orbit  \\ 
  \hline
\end{tabular}
\end{center}
\vspace{-1.5em}
\caption{List of symbols for our network model.}
\end{table}
}


\subsection{Propagation Model}
\subsubsection{Path-loss Model}
We consider the propagation effects of wireless channels by combining the path-loss attenuation and small-scale fading. For the path-loss model, we adopt the classical distance-dependent path-loss model. For example, the path-loss of a wireless channel between the $i$-th satellite in $\zeta_n$ and $\mathbf{u}$ is 
\begin{equation}
\|\mathbf{x}_{n,i}-\mathbf{u}\|^{-\alpha},
\end{equation}
where $\alpha$ is the path-loss exponent, and $\mathbf{x}_{n,i}$ is the position of the $i$-th satellite in $\zeta_n$.

We model the small-scale channel fading using the Nakagami-$m$ distribution, which captures the line-of-sight (LOS) environment of the satellite networks. Let $H_{n,i}$ be the fading coefficient of the channel between $i$-th satellite on $\zeta_n$ and $\mathbf{u}$. By assuming $\mathbb{E}[H_{n,i}]= 1$, the probability density function (PDF) of $H_{n,i}$ is given by 
\begin{equation}\label{eq:Nakagami_pdf}
f_{H_{n,i}}(x) = \frac{2m^m}{\Gamma(m)}x^{2m-1}\exp(-mx^2)\mbox{,}
\end{equation}
for $x\geq 0$ where $m$ is a shape parameter with $m\geq\frac{1}{2}$\cite{giunta2018estimation}. The Nakagami-$m$ fading model can cover a wide class of fading channel conditions. For example, when $m=1$ and $m=\frac{(K+1)^2}{2K+1}$, the fading model is reduced to the Rayleigh and Rician-$K$ distributions, respectively. By tuning the shaping parameter $m$, \eqref{eq:Nakagami_pdf} can be adjusted to fit empirical fading data sets.

\subsubsection{Beamforming Gain}
We also model the transmit and receive beamforming gain of satellites and the user. In particular, we focus on a two-lobe approximation model for the antenna radiation pattern as in \cite{bai2014coverage} which is accurate when adopting Dolph-Chebyshev beamforming weights with uniform linear arrays \cite{koretz2009dolph}. Let $G_{n,i}$ be the effective antenna gain from the $i$-th satellite in $\zeta_n$ to $\mathbf{u}$. We further assume that the user is served by one of the nearest satellites on each orbit, i.e., $\mathbf{x}_{n,1}$ for $n\in[N]$. Let $G_{\rm t}$ and $G_{\rm r}$ be the transmit and receive antenna gains for the main lobes. For analytical tractability, we assume that the receive beam of the user is aligned with $\mathbf{x}_{n,1}$ for $n\in[N]$. Whereas it is misaligned with the other interfering satellites. Therefore, the effective antenna gain at carrier frequency $f_n$ is modeled as
\begin{align}
    G_{n,1}&=G_{\rm t}G_{\rm r}\frac{c^2}{(4\pi f_n^2)} \nonumber\\
     G_{n,i}&=G_{\rm t}G_{\rm r}^{\sf SB}\frac{c^2}{(4\pi f_n^2)},\label{eq:beam_gain}
\end{align}
where $i\in \{2,\ldots, {\bar M}_n\}$  and  $c$ is the speed of light. Notice that $G_{\rm r}^{\sf SB}$ is the side-lobe beam gain of the receive antenna, which is typically less than 13 dB compared to the main beam gain ${G}_{\rm r}$, i.e., $20\log_{10}\frac{G_{\rm r}}{G_{\rm r}^{\sf SB}}\simeq {\rm 13}$dB. From this two-lobe approximation, we shall assume that the effective beam gain for the serving satellite is approximately 13 dB higher than the interfering satellites, i.e., $G_{n,1}$, which is independent of the carrier frequencies, i.e.,  $20\log_{10}\frac{G_{n,1}}{G_{n,i}}\simeq {\rm 13}$dB.

\subsection{Coverage Probability}
 We compute the coverage probability using the two-step method in \cite{park2021coverage}. We first define the coverage probability conditioned that at least one satellite exists on each arc. Let ${\bar M}_n =\Phi_n({\bar \zeta}_n)$ and $D_{n,1}$ be the number of satellites on the $n$-th orbit arc, and the distance between $\mathbf{u}$ and its nearest satellite on $\zeta_n$, respectively. In other words, $D_{n,1}=\|{\bf x}_{n,1}-{\bf u}\|_2 \leq \|{\bf x}_{n,2}-{\bf u}\|_2 \leq \cdots \leq \|{\bf x}_{n,{\bar M}_n}-{\bf u}\|_2 \leq d_{\rm max} $ for $n\in [N]$, i.e., $\{{\bf x}_{n,j}\}_{j=1}^{\bar{M}_n}\in {\bar \zeta}_n$. We further assume the user communicates with the nearest satellite in each orbit, i.e., the nearest satellite association rule. Satellites on the same orbit are assumed to share the frequency-time resources. Whereas, satellites on different orbits are assumed to use orthogonal time-frequency resources.
 

 We focus on our coverage analysis in the interference-limited regime by considering the dense satellite deployment scenario such as Starlink. The signal-to-interference ratio (SIR) for the $n$-th orbit whose geometry is parameterized by ${\bar \zeta}_n$ is given by
\begin{align}\label{eq:SINR}
{\sf SIR}_n&=\frac{ H_{n,1}\|{\bf x}_{n,1}-{\bf u}\|_2^{-\alpha} G_{n,1} P}{\sum_{i=2}^{{\bar M}_n} G_{n,i}P H_{n,i}\|{\bf x}_{n,i}-\mathbf{u}\|^{-\alpha}}=\frac{ H_{n,1}D_{n,1}^{-\alpha} }{I_n}\mbox{,}
\end{align}
where $P$ and $f_n$ denote the satellite's transmit power and the center frequency of the channel between the user and satellites on $\zeta_n$. $I_n$ is the normalized aggregated interference power given by
 \begin{align}
     I_n= \sum_{i=2}^{{\bar M}_n} \bar{G}_I H_{n,i}\|{\bf x}_{n,i}-\mathbf{u}\|^{-\alpha}\mbox{,}
 \end{align}
 where $\bar{G}_I = \frac{G_{n,i}}{G_{n,1}}$. Then, using the total probability theorem, the coverage probability can be written in terms of the important network parameters, including the satellite density $\lambda_n$, path loss exponent $\alpha$, the small-scale fading $m$, and orbit geometry parameters $\zeta_n=( R,\theta_n, R_{\mathcal{A}})$ as
\begin{align}
     P^{\sf cov}_{{\sf SIR}_n}(\gamma ;\lambda_n, \alpha, m, \zeta_n )&=\mathbb{P}\left[{\sf SIR}_n \geq \gamma\bigg{|} {\bar M}_n>0  \right]\mathbb{P}\left[  {\bar M}_n >0  \right] \nonumber\\
     &+ \mathbb{P}\left[{\sf SIR}_n \geq \gamma\bigg{|} {\bar M}_n = 0  \right]\mathbb{P}\left[  {\bar M}_n = 0  \right].\label{eq:cov1}
\end{align}
Since $\mathbb{P}\left[{\sf SIR}_n \geq \gamma\bigg{|} {\bar M}_n = 0  \right]=0$ for any positive $\gamma$, we can boil down \eqref{eq:cov1} to
\begin{align}
     P^{\sf cov}_{ {\sf SIR}_n  }(\gamma  ;\lambda_n, \alpha, m, \zeta_n  )&=\mathbb{P}\left[{\sf SIR}_n \geq \gamma\bigg{|} {\bar M}_n >0  \right]\mathbb{P}\left[  {\bar M}_n>0  \right]. \label{eq:cov2}
\end{align}
As a result, to compute the coverage probability, we need to first compute the conditional coverage probability
\begin{align}
    P^{\sf cov}_{ {\sf SIR}_n|{\bar M}_n >0 }(\gamma  ;\lambda_n, \alpha, m, \zeta_n )=\mathbb{P}\left[{\sf SIR}_n \geq \gamma\bigg{|} {\bar M}_n >0  \right]
\end{align}
and then marginalize it with respective to $\mathbb{P}\left[  {\bar M}_n>0  \right]$.


We also define the coverage probability when a receiver can communicate with satellites placed on different orbits. In particular, we consider an opportunistic communication scenario, in which a receiver selects the satellite that yields the maximum instantaneous SIR when satellites are placed at different orbits. Under the premise that satellites on different orbits use the orthogonal time-frequency resources, the coverage probability can be defined as
\begin{align}\label{eq:max_SIR}
P^{\sf cov}_{{\sf SIR}_{\rm max}}\left(\gamma; \{\lambda_n\}_{n=1}^{N}, \alpha, m, \{\zeta_n\}_{n=1}^{N}\right) = \mathbb{P}\left[\max_{n\in [N]}{\sf SIR}_n \geq \gamma\right].
\end{align}
This coverage probability extends the coverage probability in \eqref{eq:cov2} in the sense of exploiting multiple orbits for opportunistic communications. In Section \ref{sec:multiple_orbits}, we will explore how the coverage probability can improve by increasing the number of orbits. 

It is worth mentioning that the coverage probability differs from the coverage probability seen by a typical user as in \cite{park2021coverage}. This coverage probability in \eqref{eq:cov2} varies as a function of the user's location and orbit geometries. Consequently, it can suitably capture the user's location and orbit geometry-specific performance. 

\section{Satellite Visible Probability}\label{sec: Section3}
In this section, we characterize the probability that at least one satellite is visible to the user. This visible probability is highly dependent on both users' location and satellite geometry parameters, $\bar{\zeta}_n=(R, \theta_n, R_{\mathcal{A}})$ where i) the Polar angle of the satellite orbit $\theta_n$, ii) the radius of the satellite orbit $R$, and iii) the distance between the center of the Earth and the base of $\mathcal{A}$, $R_{\mathcal{A}}=d_{\rm max}\sin \omega_{\rm min} +R_{\sf E}$. For ease of exposition, we assume that the user's location is fixed at ${\bf u}=(0,0,R_{\sf E})$ in the Cartesian coordinate system and the satellite parameters are chosen for arbitrary values. As a stepping stone toward computing the visibility probability, we first derive the length of the satellites' visible trajectory in terms of the satellite's orbit parameter $\bar{\zeta}_n$, which is stated in the following lemma.

\lemma\label{lem1}  The length of the satellites' visible trajectory on the orbit $\bar{\zeta}_n=(R, \theta_n, R_{\mathcal{A}})$ is given by 
\begin{equation}\label{eq:lemma1}
    L(R,\theta_n,R_{\mathcal{A}}) =\begin{cases}
 R \arccos{\left(\eta(R,\theta_n,R_{\mathcal{A}})\right)} & \mbox{for }|\theta_n-\frac{\pi}{2}| \leq \arccos\left(\frac{R_{\mathcal{A}}}{R}\right)\\
0 &  \mbox{otherwise} 
\end{cases}\mbox{,}
\end{equation}
where 
\begin{align}\label{eq:kappa}
    \eta(R,\theta_n,R_{\mathcal{A}})= {\frac{2R_{\mathcal{A}}^2}{R^2\sin^2{\theta_n}} -1}\mbox{.}
\end{align}
\begin{IEEEproof}
See Appendix \ref{appen:lengthonvisiblesurface}.
\end{IEEEproof}

\begin{figure}[t]
\begin{center}
\includegraphics[scale=0.75]{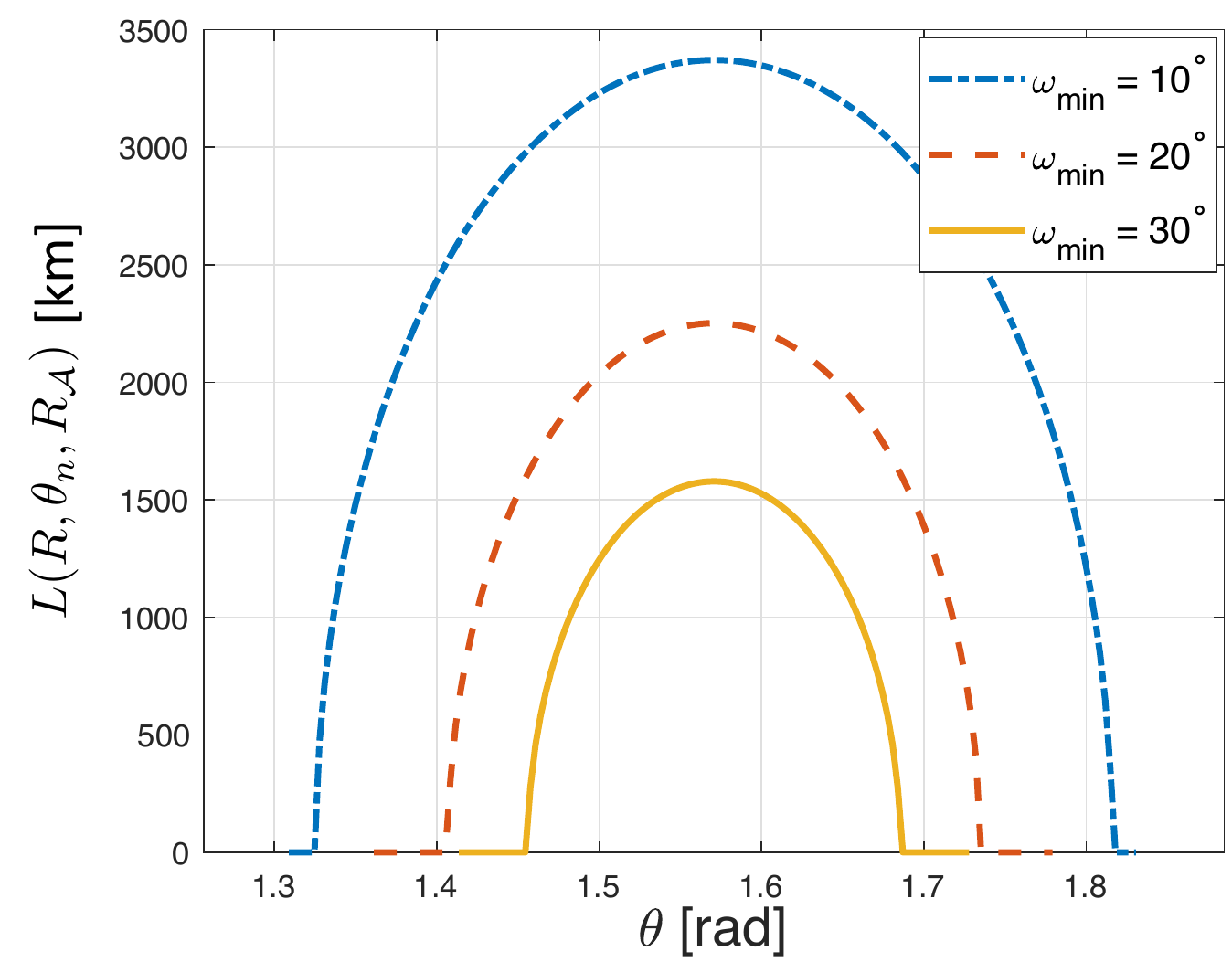}
\end{center}
\vspace{-0.5em} 
\caption{Length of visible orbit trajectory according to $\omega_{\rm min}\in \{10^{^{\circ}},20^{^{\circ}},30^{^{\circ}}\}$ when $R_{\sf E} =637$km and $R_{\sf h}=500$km.} 
\label{fig:net_top} 
\end{figure}

\begin{figure}[t]
\begin{center}
\includegraphics[scale=0.73]{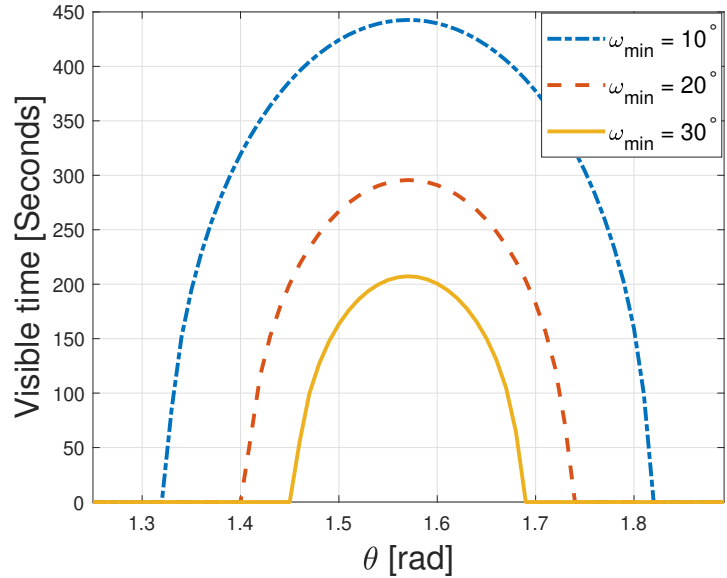}
\end{center}
\vspace{-0.5em} 
\caption{Visible time according to $\omega_{\rm min}\in \{10^{^{\circ}},20^{^{\circ}},30^{^{\circ}}\}$ when $R_{\sf E} =6371 (km)$ and $R_{\sf h}=500(km)$.} 
\label{fig:time} 
\end{figure}

We provide the implication for Lemma 1. As in \eqref{eq:lemma1}, the length of the satellites' visible trajectory on the orbit is determined by two factors, the radius of the satellite's orbit $R$ and the arccos term $\arccos(\eta(R,\theta_n,R_{\mathcal{A}}))$. The geographical meaning of the argument of the {\rm arccos} term, $\eta(R,\theta_n,R_{\mathcal{A}})$, is the vertex angle of the isosceles triangle whose vertices are the center of the Earth and the intersecting points of the orbit with the polar angle $\theta_n$ and the plane $R_{\mathcal{A}}$. To provide a more clear understating, we depict the length of the satellite's visible trajectory as a function of the polar angle $\theta_n$ for three different minimum elevation angles $\omega_{\rm min}\in \{10^{^{\circ}},20^{^{\circ}},30^{^{\circ}} \}$, which can vary $R_{\mathcal{A}}=d_{\rm max}\sin \omega_{\rm min} +R_{\sf E}$. As illustrated in Fig. \ref{fig:time}, the length has the maximum when the polar angle is $\theta_n=\frac{\pi}{2}$ for all $\omega_{\rm min}$. This confirms our intuition that the satellites are more visible when the normal unit-vector $\mathbf{v}_n =(1,\theta_n,\phi_n)$ of the orbit $\zeta_n$ is perpendicular to the user's location vector ${\bf u}=(0,0,R_{\sf E})$. Therefore, when the user's location is changed, the maximum polar angle can be different. Another interesting point is that the length of the satellites' visible trajectory is inversely proportional to minimum elevation angles, i.e., the length increases as the minimum elevation angle reduces. This also aligned with our intuition that the satellite is more visible when the visible spherical cap is larger.

{\bf Example 1:} Consider some special cases better to understand the length of the satellite's visible trajectory. The special case is when the user can observe the maximum length of $\bar{\zeta}_n$ when $\theta_n = \frac{\pi}{2}$, i.e., the orbit passes the zenith of $\mathbf{u}$. In this case, the length in \eqref{eq:lemma1} simplifies to
\begin{equation}\label{eq:example1}
    L\left(R,\frac{\pi}{2},R_{\mathcal{A}}\right)=R \arccos\left( 2 \frac{R_{\mathcal{A}}^2}{R^2}-1\right)\mbox{.}
\end{equation}
When $\omega_{\rm min}= 0$, i.e., $R_{\mathcal{A}}=R_{\sf E}$, the length is further simplified to $L\left(R,\frac{\pi}{2},R_{\sf E}\right) = R\arccos\left(2\frac{R_{\sf E}^2}{R^2}-1\right)$.

\vspace{0.2cm}

 From Lemma \ref{lem1}, we establish the satellite visibility probability when satellites are placed according to a homogeneous PPP with an intensity $\lambda$ on the orbit ${\zeta}_n$.

\lemma \label{lem2}
Suppose satellites are distributed on the orbit $\zeta_n$ for $n\in [N]$ according to independent homogeneous PPP with intensity $\lambda$. Then, the probability that at least one satellite is visible to the user is
\begin{align}
    P_{\sf visible}(\lambda, \{\zeta_n\}_{n\in [N]})= 1-\exp\left(- \lambda \sum_{n=1}^N L(R,\theta_n,R_{\mathcal{A}})\right).\label{eq:lem2}
\end{align}

\begin{IEEEproof}
 The proof is direct from the void probability for the homogeneous PPP. Since we assume that the satellites are distributed as the homogeneous PPP with the intensity $\lambda$ on the orbit ${\zeta}_n$, the mean number of satellites onto the visible orbit trajectory with length $L(R,\theta_n,R_{\mathcal{A}})$ is $\lambda L(R,\theta_n,R_{\mathcal{A}})$. Further, we know that the sum of independent Poisson random variables with mean $\lambda L(R,\theta_n,R_{\mathcal{A}})$ forms another Poisson random variable with mean $\lambda\sum_{n=1}^N L(R,\theta_n,R_{\mathcal{A}})$. Therefore, the probability that at least one satellite exists on the visible trajectory is given by \eqref{eq:lem2}. This completes the proof.
 \end{IEEEproof}

{\bf Remark 1 (Connection to a Cox point process):} It is worth mentioning that the locations of visible satellites do not form a homogeneous PPP anymore on the surface of the spherical cap. This is because the union of the independent PPPs on the lines, i.e., the on-dimensional space, is not PPP on the surface of the spherical cap, i.e., the two-dimensional space. If the polar angles of $N$ orbits, $\{\theta_n\}_{n\in[N]}$, follow the Poisson distribution, the satellites on $\mathcal{A}$, i.e., $\left(\bigcup_{n=1}^N\Phi_n\right)\cap\mathcal{A}$, form a Cox point process, i.e., a doubly stochastic process, as defined in \cite{baccelli2009stochastic}. Modeling the locations of the satellites according to a Cox process is not the scope of this paper, but it would be an interesting research direction for the coverage analysis of satellite networks.

%

{\bf Remark 2 (Satellite visible time):} It is also interesting to compute the time to communicate with the satellites onto orbit geometry $\zeta_n$. Let $G(=6.67259\times 10^{-11} (m^3 kg^{-1}s^{-2}))$ and $M(=5.9736\times 10^{24}(kg))$ be the universal constant of gravitation and the mass of the Earth, respectively. From the balance condition between the centrifugal force and the gravity, the speed of the LEO satellite is given by $\sqrt{\frac{GM}{R}}$. Using the speed and the visible length $L(R,\theta_n,R_{\mathcal{A}})$ in Lemma 1, the visible time of satellite passing $\zeta_n$ by the user located at ${\bf u}$ is 
\begin{align}
    \tau_{\sf visible}(\zeta_n) = \frac{L(R,\theta,R_{\mathcal{A}})\sqrt{R}}{\sqrt{GM}}
\end{align}
in seconds. For instance, if we put $R_{\sf E} = 6371 (km)$, $R_{\sf h} = 500 (km)$, $\omega_{\rm min} = 10^{\circ}$, and $\theta = \frac{\pi}{2}$  (rad), the satellite visible time is 442.6396 seconds where $L(R,\frac{\pi}{2},R_{\mathcal{A}}) = 3.3714 \times 10^6$ meters and the satellite velocity is $7.6165\times 10^3(m/s)$. The visible times with different $\omega_{\rm min}$ under $R_{\sf E} =6371 (km)$ and $R_{\sf h}=500(km)$ are given in Figure \ref{fig:time}.

\section{Coverage Analysis for Single-Orbit Networks}\label{sec: Section4}
This section provides exact expressions of the coverage probability for a single-orbit satellite network with orbit geometry $\zeta_n$. To compute the coverage probability, we first introduce two important lemmas. The former characterizes the nearest distance distribution for the satellite when at least one satellite exists on the visible trajectory ${\bar \zeta_n}$. The next one is the Laplace transform of the aggregated interference power conditioning that the nearest satellite distance is fixed. Using these two lemmas as building blocks, we will establish the coverage probability expression in the sequel.




\vspace{0.3cm}
\lemma\label{lem:neares_satellite_1} 
The CCDF of $D_{n,1}$ conditioned on that at least one satellite exists on ${\bar \zeta}_n$, i.e., $\bar{M}_n>0$ is  
\begin{align}\label{eq:ccdf_t1}
    F^c_{D_{n,1}|\bar{M}_n>0}(r)=\frac{\exp\left(-\lambda R\arccos\left(\eta(R,\theta_n,\frac{R^2+R_{\sf E}^2-r^2}{2R_{\sf E}})\right)\right)-\exp(-\lambda L(R,\theta_n,R_{\mathcal{A}}))}{1-\exp(-\lambda L(R,\theta_n,R_{\mathcal{A}}))}\mbox{,}
\end{align}
and the PDF of $D_{n,1}$ conditioned on $\bar{M}_n>0$ is
\begin{align}\label{eq:pdf_t1}
    f_{D_{n,1}|\bar{M}_n>0} (r) =\frac{2r\lambda(-r^2+R^2+R_{\sf E}^2)\exp\left(-\lambda R \arccos\left(\eta(R,\theta_n,\frac{R^2+R_{\sf E}^2-r^2}{2R_{\sf E}})\right)\right)}{RR_{\sf E}^2\sin^2{\theta_n}(1-\exp(-\lambda L(R,\theta_n,R_{\mathcal{A}})))\sqrt{1-\left(\eta(R,\theta_n,\frac{R^2+R_{\sf E}^2-r^2}{2R_{\sf E}})\right)^2}}
\mbox{,}
\end{align}
for $d_{\rm min}(\theta_n)<r<d_{\rm max}$ where $d_{\rm min}(\theta) = \sqrt{R^2-2R_{\sf E} R\sin(\theta)+R_{\sf E}^2}$ is the minimum distance between $\mathbf{u}$ and $\zeta_n$.
\begin{IEEEproof}
See Appendix. \ref{appen:nearest_satellite_1}
\end{IEEEproof}
 
 \vspace{0.2cm}
 Unlike the prior work in \cite{park2021coverage, andrews2011tractable}, the nearest distance distribution derived in Lemma \ref{lem:neares_satellite_1} is unwieldy to interpret because of the complicated orbit geometry parameters. Notwithstanding, we can observe that the CCDF is mainly determined by two probability terms: i) $\exp\left(-\lambda R \arccos\left(\eta(R,\theta_1,\frac{R^2+R_{\sf E}^2-r^2}{2R_{\sf E}})\right)\right)$ and ii) $1-\exp(-\lambda L(R,\theta_n,R_{\mathcal{A}}))$. The former one is the probability that no satellite exists on the arc of $\zeta_n$ whose distance from $\mathbf{u}$ is smaller than $r$. The later one is the probability that there exists at least one visible satellite at $\mathbf{u}$ on $\bar{\zeta}_n$, i.e., $\mathbb{P}[\bar{M}_n>0]$. The ratio of these two void probabilities determines the CCDF of the conditional nearest distance distribution. 
 
\begin{figure}[t]
\begin{center}
\includegraphics[scale=0.75]{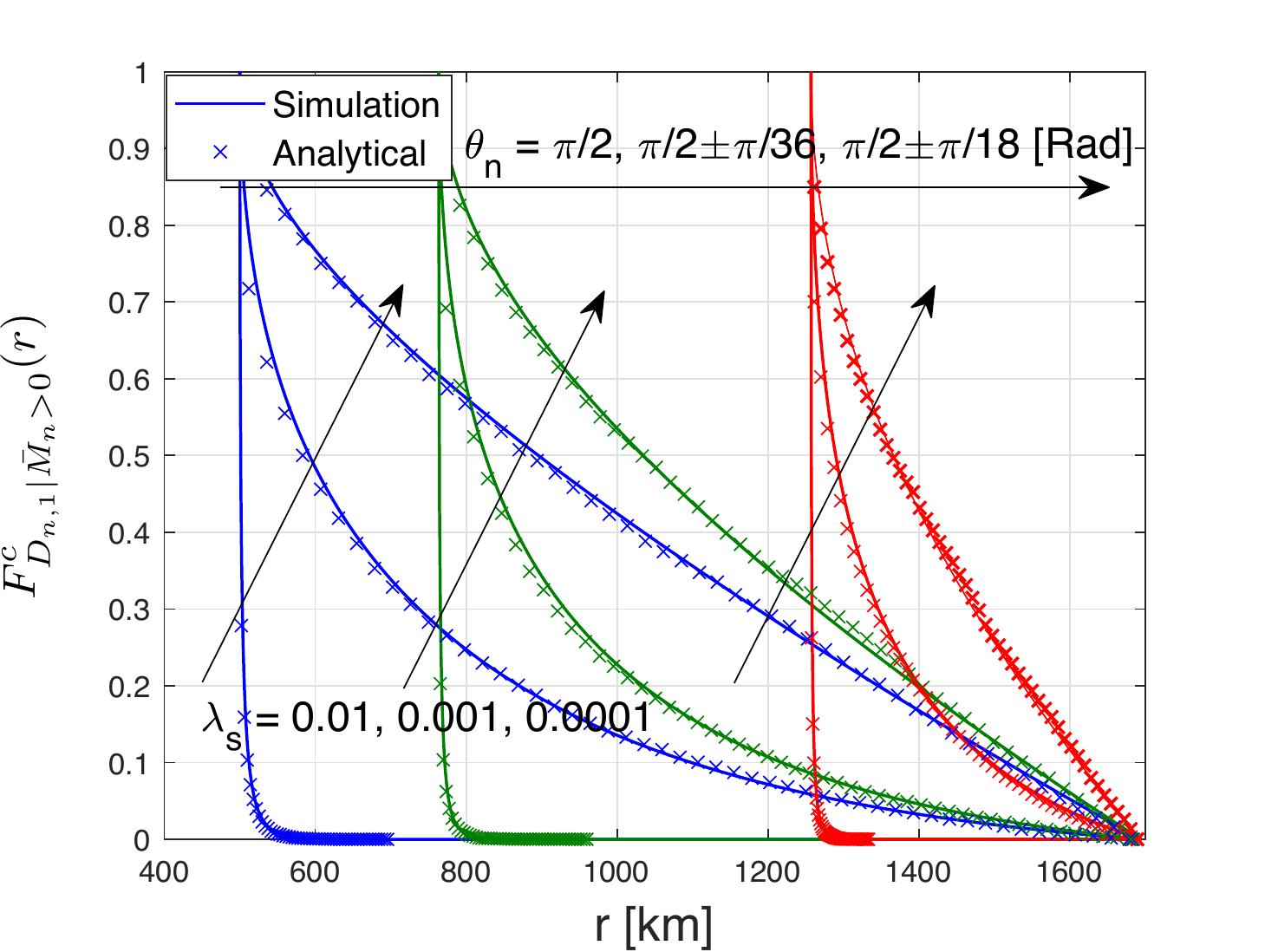}
\end{center}
\vspace{-0.5em} 
\caption{The conditional CCDF of the nearest distance for various parameter values of $ \theta_n\in \{\frac{\pi}{2}, \frac{\pi}{2}\pm\frac{\pi}{36},\frac{\pi}{2}\pm \frac{\pi}{18}\}$ and $\lambda \in \{0.01, 0.001, 0.0001\}$ for fixed $R_{\sf E} = 6371, R_{\sf h} = 500,$ and $ \omega_{\rm min} = 10^{\circ}$.} 
\label{fig:near_dist_cdf} 
\end{figure}

 To provide a more precise understanding and validate our analytical expression in Lemma \ref{lem:neares_satellite_1}, as shown in Figure \ref{fig:near_dist_cdf}, we illustrate the CCDFs for various orbit geometry parameters, $\theta_n \in \{\frac{\pi}{2}, \frac{\pi}{2}\pm\frac{\pi}{36},\frac{\pi}{2}\pm \frac{\pi}{18}\}$ and $\lambda \in \{ 0.01, 0.001, 0.0001\}$ for fixed $R_{\sf E} = 6371$, $R_{\sf h} = 500$, and $\omega_{\rm min} = 10^{\circ}$. From the simulations, we observe that the nearest distance $D_{n,1}$ increases as the orbit becomes more tilted. In addition, the CCDF of the nearest distance becomes more heavy-tailed as the density of satellites decreases, which also agrees with our intuition. Therefore, we can conclude that the nearest satellite is more accessible when $\theta_n =\frac{\pi}{2}$ and the density is high. Furthermore, by comparing our analytical expression with the simulations, we verify that our analysis is exact. 

Next, we introduce the lemma for the Laplace transform of the aggregated interference power.

\lemma\label{lem:Laplace_interference} Conditioned that the nearest satellite is located with the distance of $D_{n,1} = r$, the Laplace transform of the aggregated interference power from outside of the nearest distance is given by
\begin{align}\label{eq:laplace}
    &\mathcal{L}_{I_n|D_{n,1}=r}(s) =\exp\bigg{(}-\frac{\lambda}{RR_{\sf E}^2\sin^2(\theta_n)}\times
    \nonumber\\
    &\int_{r}^{{d_{\rm max}}}\left[ 1 -\left(1+\frac{s\bar{G}_I {u}^{-\alpha}}{m}\right)^{-m}\right] \frac{2u(-u^2+R^2+R_{\sf E}^2)}{ \sqrt{1-\left(\eta\left(R,\theta_n,\frac{-u^2+R^2+R_{\sf E}^2}{2R_{\sf E}}\right)\right)^2}} {\rm d}u                 \bigg{)}\mbox{,}      
\end{align}
for $d_{\rm min}(\theta_n)<r<d_{\rm max}$.

\begin{IEEEproof}
See Appendix \ref{appen:Laplace_interference}.
\end{IEEEproof}

\subsection{SIR Coverage Analysis}

Now, we are ready to present our main result for the coverage probability, which is stated in the following theorem. 

\theorem\label{theo:cov_prob} Conditioned that at least one satellite exists on the visible trajectory ${\bar \zeta}_n$, i.e., ${\bar M}_n>0$, the SIR coverage probability is given by
\begin{align}\label{eq:cov_prob} 
   P^{\sf cov}_{ {\sf SIR}_n|{\bar M}_n >0 }(\gamma  ;\lambda_n, \alpha, m, \zeta_n )
 = \int_{d_{\rm min}(\theta_n)}^{d_{\rm max}}  
   \sum_{t=0}^{m-1}\frac{(-m r^{\alpha })^t}{t!} \frac{d^t\mathcal{L}_{I_n|D_{n,1}=r}(s)}{ds^t}\bigg{|}_{s =m\gamma r^{\alpha}}
   f_{D_{n,1}|\bar{M}_n>0} (r)    {\rm d}r \mbox{,}
\end{align}
and the coverage probability is
\begin{equation}\label{eq:cov_prob_prob}
    P^{\sf cov}_{ {\sf SIR}_n }(\gamma  ;\lambda_n, \alpha, m, \zeta_n )
 = P^{\sf cov}_{ {\sf SIR}_n|{\bar M}_n >0 }(\gamma  ;\lambda_n, \alpha, m, \zeta_n )
\left(1-\exp(-\lambda L(R,\theta_n,R_{\mathcal{A}}))\right)\mbox{.}
\end{equation}
\begin{IEEEproof}
See Appendix \ref{appen:cov_prob}.
\end{IEEEproof}

The SIR coverage probability expression in Theorem 1 has full-generality in the sense of having all relevant parameters, including fading effect $m$, path-loss exponent $\alpha$, the density $\lambda_n$, and orbit geometry $\zeta_n$. This analytical expression, however, has the lack of tractability because it involves multiple integrals and derivatives of the complicated functions. For a better understanding, we provide some simulation results for the SIR coverage probability by tuning the relevant network parameters. Furthermore, we confirm the exactness of our analysis by comparing it with simulation results for various parameters.



\begin{figure}[t]
\begin{center}
\includegraphics[scale=0.75]{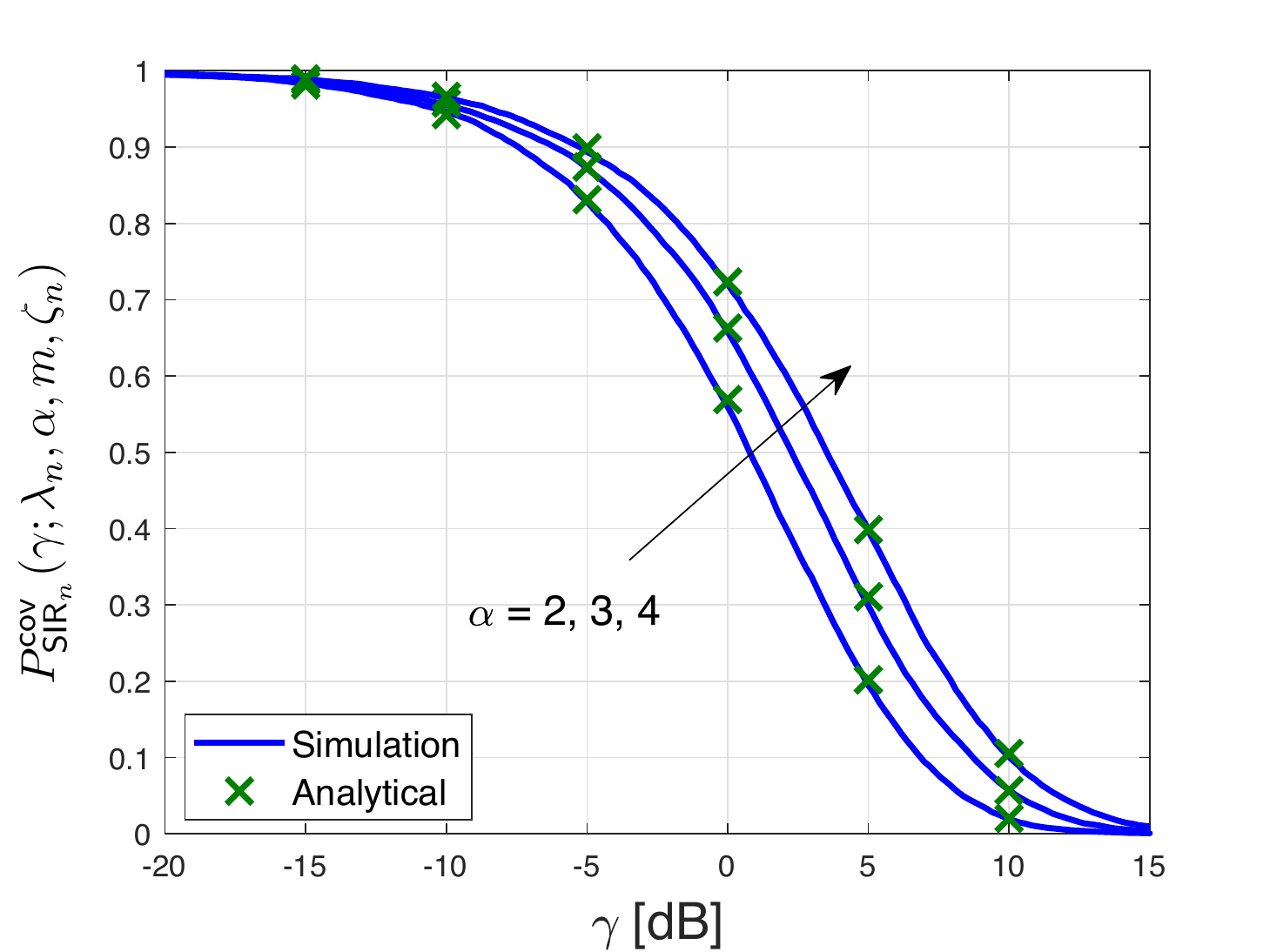}
\end{center}
\vspace{-0.5em} 
\caption{Effects of path-loss exponent $\alpha$ on the coverage probability for fixed $R_{\sf h} = 500$, $\omega_{\rm min}= 10^{\circ}$, $\lambda = 0.005$, and $m=1$. }
\label{fig:effect_ch_exp} 
\end{figure}

\textbf{Effect of pass-loss exponent:} The interference power decays faster than the serving signal power as $\alpha$ increases since the association rule is the nearest satellite association. Therefore, we can expect an enhanced coverage probability for a larger channel exponent. Figure \ref{fig:effect_ch_exp} shows this improvement by increasing $\alpha$ from 2, 3 and 4 for fixed $R_{\sf h} = 500$, $\omega_{\rm min}= 10^{\circ}$, $\lambda = 0.005$, and $m=1$.

\begin{figure}[t]
\begin{center}
\includegraphics[scale=0.75]{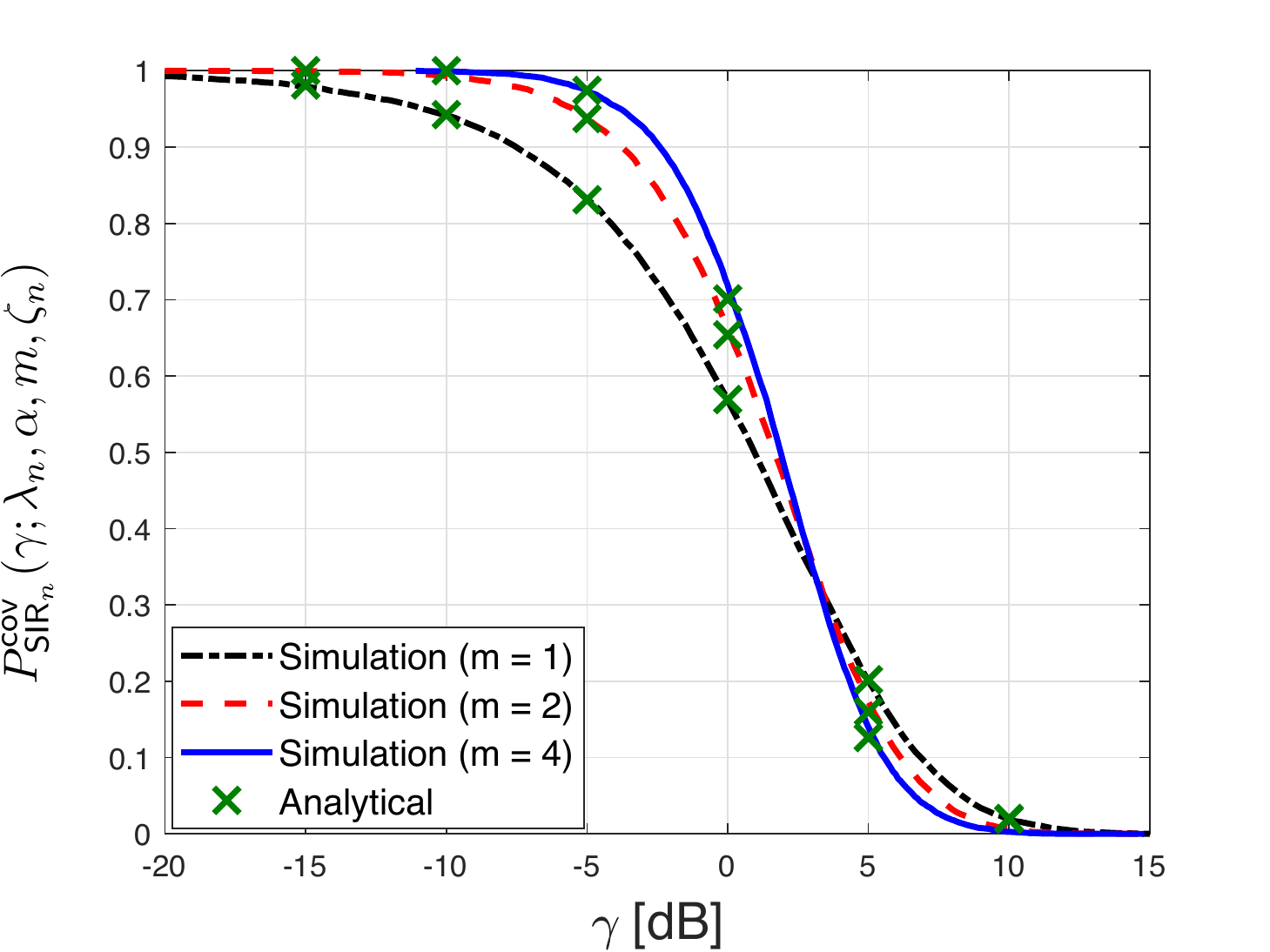}
\end{center}
\vspace{-0.5em} 
\caption{Effects of Nakagami fading parameter $m$ on the coverage probability for fixed $R_{\sf h} = 500$, $\omega_{\rm min}= 10^{\circ}$, $\lambda = 0.005$, and $\alpha=2$.  }
\label{fig:effect_mu} 
\end{figure}

\textbf{Effects of the LOS strength in fading:} Figure \ref{fig:effect_mu} shows the coverage probability for different values of $m$, which is the shape parameter of the Nakagami distribution for fixed $R_{\sf h} = 500$, $\omega_{\rm min}= 10^{\circ}$, $\lambda = 0.005$, and $\alpha=2$. In the Nakagami-$m$ distribution, the average fading power becomes small as $m$ decreases, but the variance increases. Therefore, a larger $m$ enhances the coverage probability by serving the higher average power in a low SIR regime. Whereas, in a high SIR regime, a smaller $m$ is beneficial to increase the coverage probability due to its higher variance.

\begin{figure}[t]
\begin{center}
\includegraphics[scale=0.75]{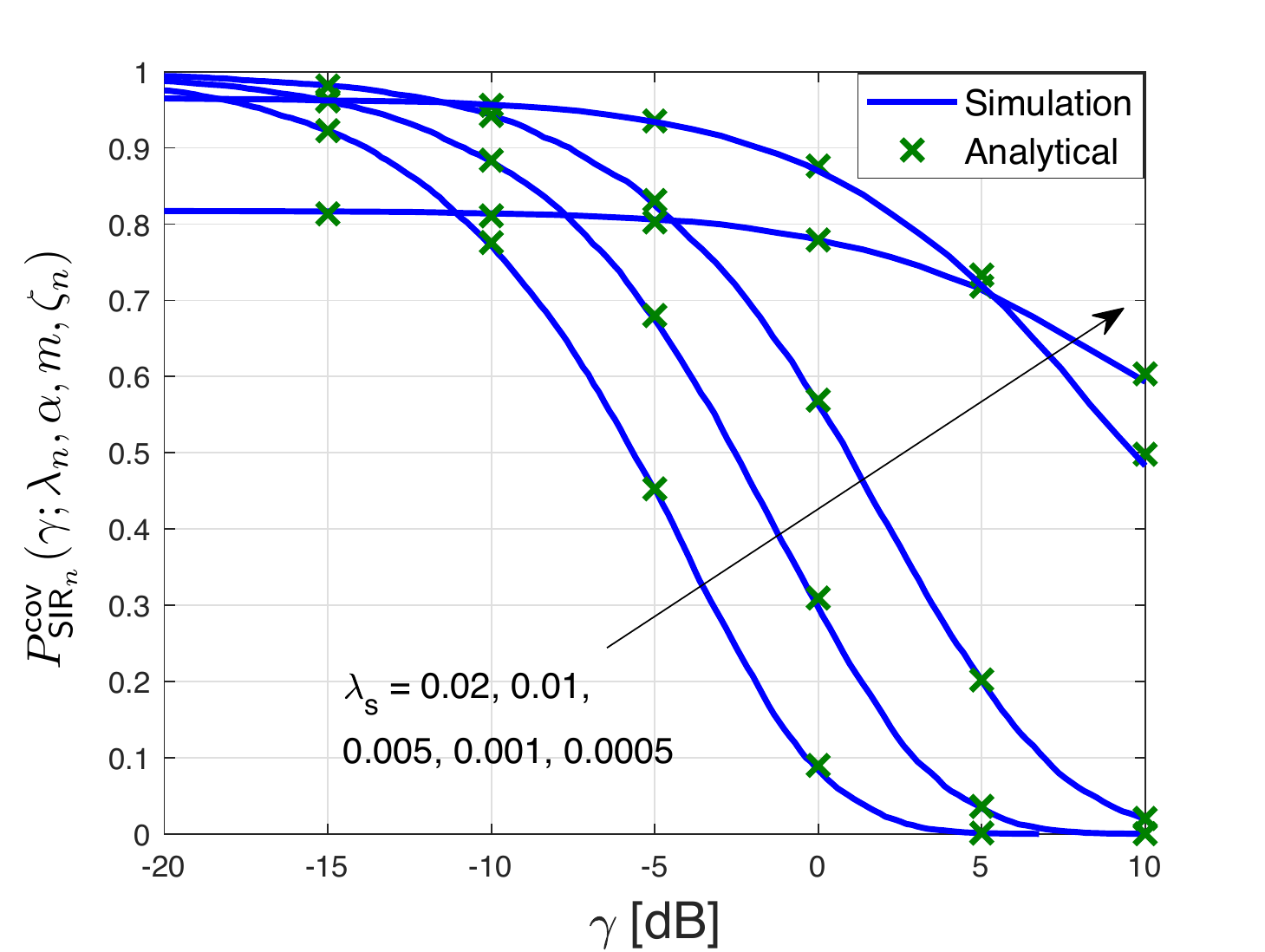}
\end{center}
\vspace{-0.5em} 
\caption{Effects of satellite density $\lambda$ on the coverage probability. $R_{\sf h} = 500$, $\omega_{\rm min}=10^{\circ}$, $\theta_n=\frac{\pi}{2}$, $\alpha=2$, and $m = 1$.}
\label{fig:effect_density} 
\end{figure}

\textbf{Effects of the network density:} We illustrate the effects of $\lambda$ on the coverage probability in Figure \ref{fig:effect_density} for fixed $R_{\sf h} = 500$, $\omega_{\rm min}=10^{\circ}$, $\theta_n=\frac{\pi}{2}$, $\alpha=2$, and $m = 1$. We can observe that the coverage probability improves as $\lambda$ decreases for the high SIR region. Unlike the conventional cellular network model in terrestrial networks \cite{andrews2011tractable}, $D_{n,1}$ is lower-bounded by $d_{\min}(\theta_n) = \sqrt{R^2-2R_{\sf E}R\sin(\theta_{n})+R_{\sf E}^2}$ in our satellite network model. As a result, it leads to the performance degradation as the density $\lambda$ increases. Nevertheless, this performance tendency can be broken down when the density is extremely small. In this case, the probability of no visible satellite on $\bar{\zeta}_n$ becomes high, and the user cannot be served by any satellites on $\zeta_n$. So, it leads to degrading the coverage performance severely. This phenomenon can be observed in Figure \ref{fig:effect_density} when $\lambda$ is chosen as 0.0005.

\begin{figure}[t]
\begin{center}
\includegraphics[scale=0.75]{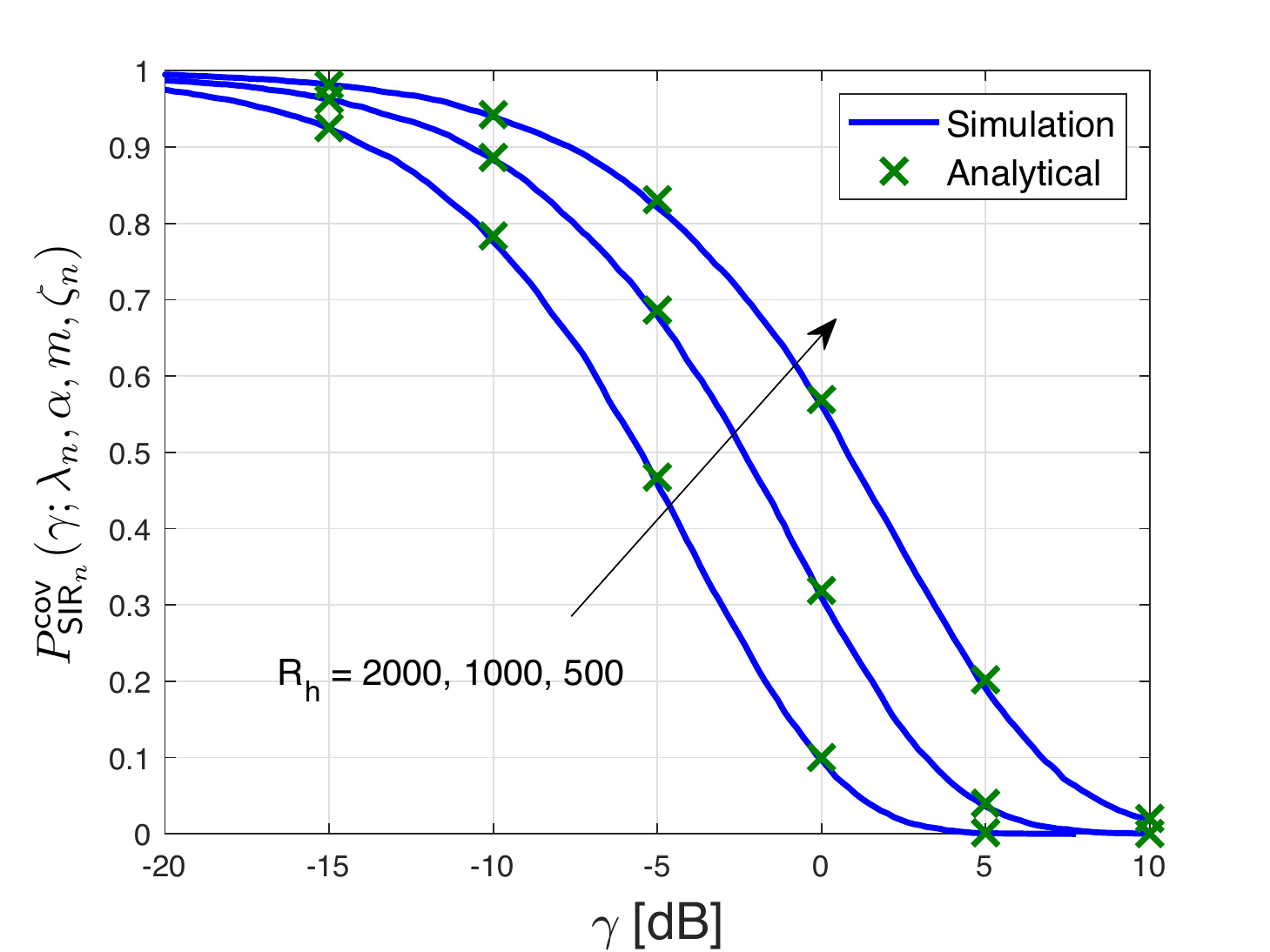}
\end{center}
\vspace{-0.5em} 
\caption{Effects of satellite's altitude $R_{\sf h}$ on the coverage probability. $\lambda = 0.005$, $\omega_{\rm min}=10^{\circ}$, $\theta_n=\frac{\pi}{2}$, $\alpha=2$, and $m = 1$. }
\label{fig:effect_altitude} 
\end{figure}

\textbf{Effects of the satellite's altitude:} In Figure \ref{fig:effect_altitude}, for fixed $\lambda = 0.005$, $\omega_{\rm min}=10^{\circ}$, $\theta_n=\frac{\pi}{2}$, $\alpha=2$, and $m = 1$. By increasing $R_{\sf h}$, we examine how the coverage probability is affected by the satellite's altitude $R_{\sf h}$. As can be seen in this figure, the coverage performance improves as decreasing the altitude of the satellite orbit. This is mainly because the mean number of interfering satellites is proportional to the length of visible trajectory $L(R,\theta,R_{\mathcal{A}})$.

\begin{figure}[t]
\begin{center}
\includegraphics[scale=0.75]{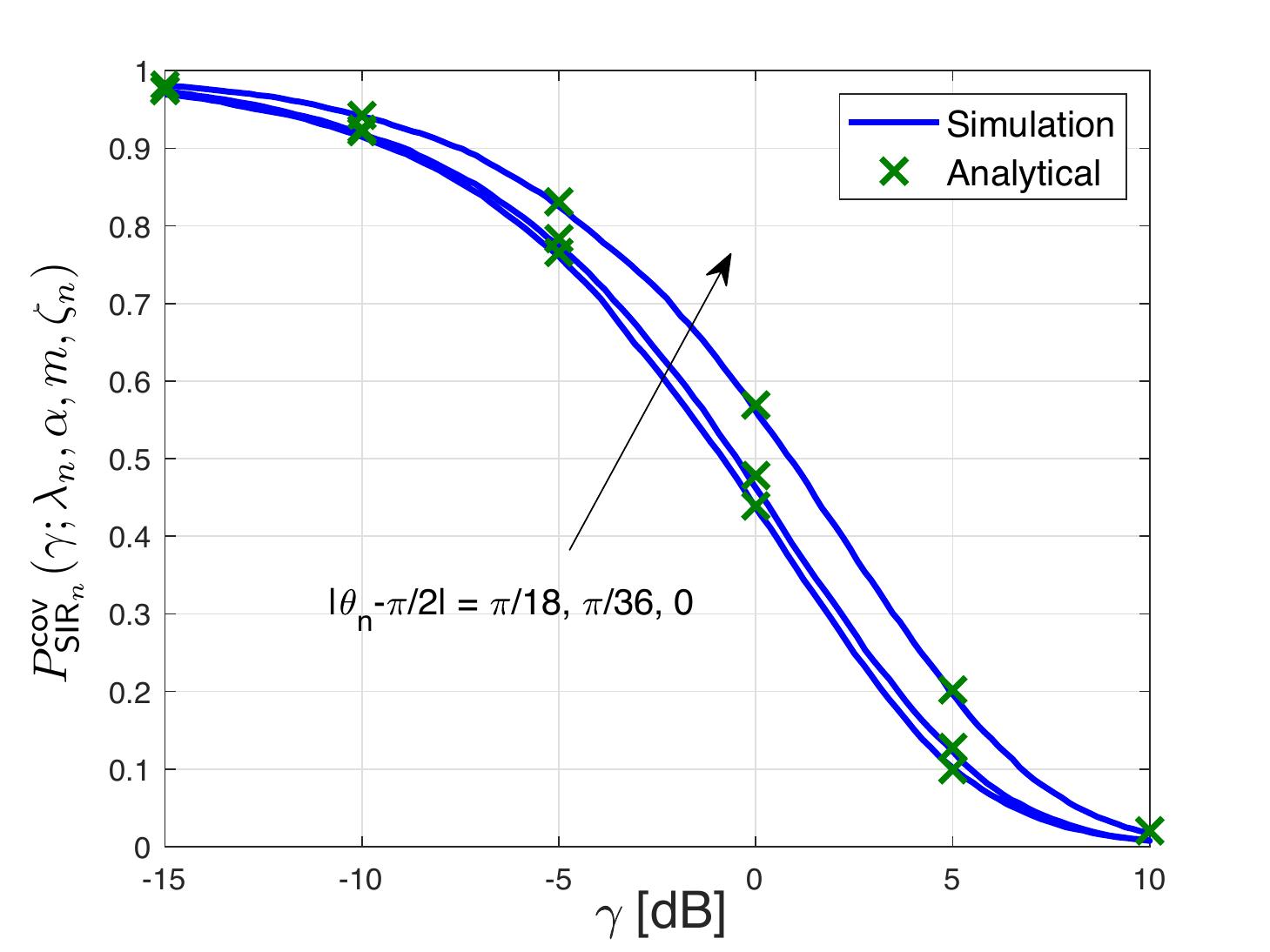}
\end{center}
\vspace{-0.5em} 
\caption{Effects of orbit angle $\theta_n$ on the coverage probability. $\lambda = 0.005$, $\omega_{\rm min}=10^{\circ}$, $R_{\sf h} = 500$, $\alpha=2$, and $m = 1$.  }
\label{fig:effect_theta} 
\end{figure}

\textbf{Effects of the orbit angle:} Figure \ref{fig:effect_theta} shows the effects of orbit polar angle $\theta_n$ on the coverage probability for fixed $\lambda = 0.005$, $\omega_{\rm min}=10^{\circ}$, $R_{\sf h} = 500$, $\alpha=2$, and $m = 1$. As expected, we can observe that the coverage performance enhances as the orbit is more tilted because of a better nearest distance distribution for the serving satellite. Especially, when $\zeta_n$ passes the zenith of the user, the highest coverage performance is achievable.


\begin{figure}[t]
\begin{center}
\includegraphics[scale=0.75]{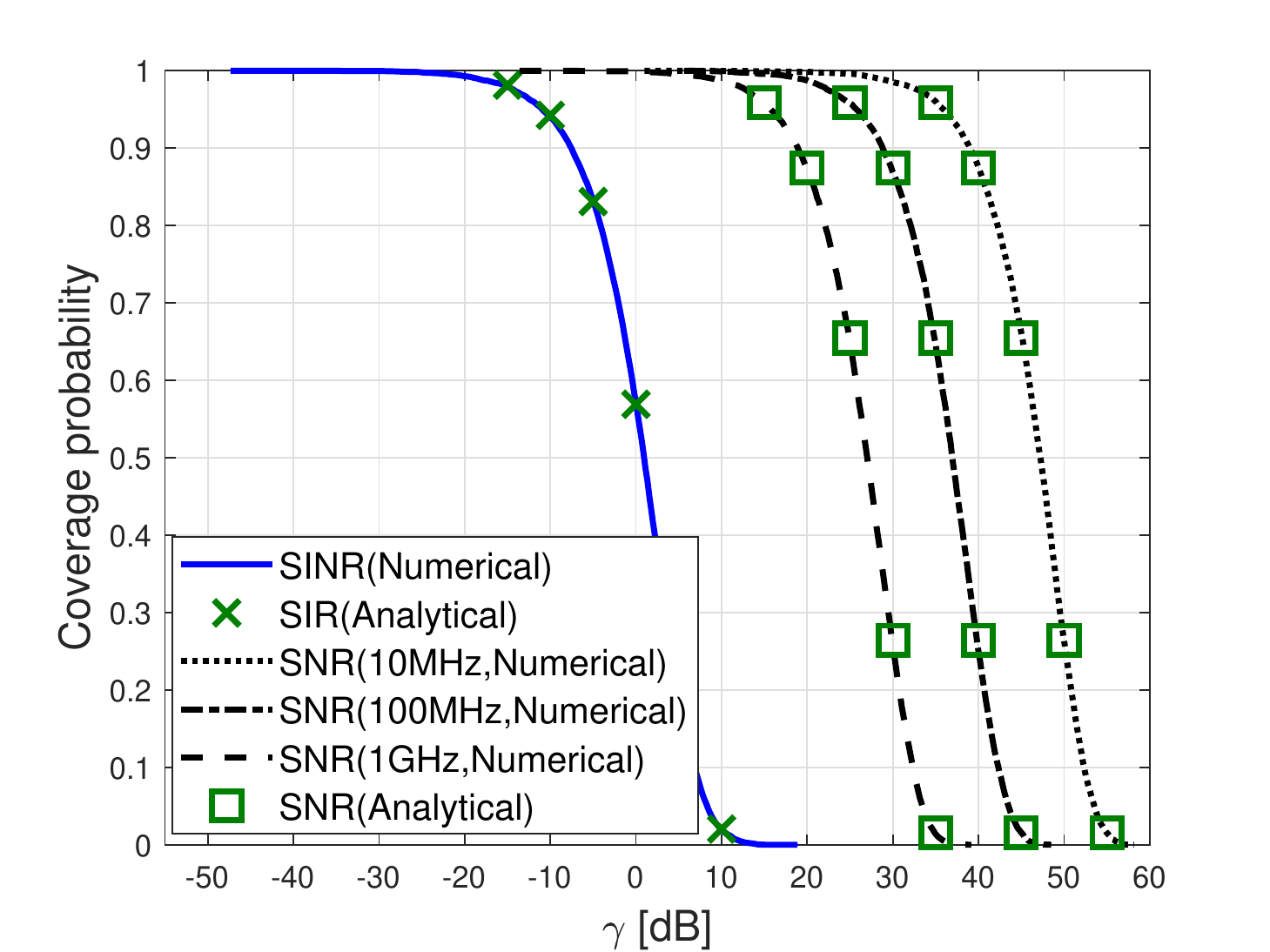}
\end{center}
\vspace{-0.5em} 
\caption{Comparison of SINR, SIR, SNR-based coverage probabilities according to the signal bandwidth of 10, 100, and1000MHz. In this simulation, we set $R_h=500$, $\omega_{min}=10^{\circ}$, $\theta_n=\frac{\pi}{2}$, $\lambda = 0.005$, $P$ = 40 dBm, and $G_{n,1}$=30 dBi.  }
\label{fig:SNR} 
\end{figure}

\subsection{SNR Coverage Probability}

We have performed the SIR coverage analysis by capitalizing on the interference-limited regime. However, when the satellites in the same orbit use orthogonal time-frequency resources mutually, the coverage performance can be mainly determined by the signal-to-noise ratio (SNR) instead of SIR. In this subsection, we extend our coverage analysis for the SNR metric. We commence by defining SNR for the $n$-th orbit as 
\begin{align}
{\sf SNR}_n &=\frac{ H_{n,1}\|{\bf x}_{n,1}-{\bf u}\|_2^{-\alpha} }{\frac{\sigma^2}{PG_{n,1}}}\mbox{,}
\end{align}
where $\sigma^2$ is the noise power. The following Corollary states the SNR coverage performance for the single-orbit satellite network. 


\corollary \label{cor1} In the noise-limited regime, the SNR coverage probability conditioned on $\bar{M}_n>0$ is given by
\begin{align}\label{eq:cov_prob_SNR} 
   &P^{\sf cov}_{{\sf SNR}_n|\bar{M}_n>0}(\gamma ;\lambda_n, \alpha, m, \zeta_n ) =\nonumber\\
   &\int_{d_{\rm min}(\theta_n)}^{d_{\rm max}}  
   \exp\left(-m\frac{\sigma^2}{PG_{n,1}}\gamma r^{\alpha}\right)\sum_{t=0}^{m-1}\frac{\left(m\sigma^2\gamma r^{\alpha}\right)^t}{(PG_{n,1})^t t!}
   f_{D_{n,1}|\bar{M}_n>0} (r)    dr \mbox{,}
\end{align}
and the coverage probability is
\begin{equation}\label{eq:cov_prob_prob_SNR}
    P^{\sf cov}_{{\sf SNR}_n}(\gamma ;\lambda_n, \alpha, m, \zeta_n ) = P^{\sf cov}_{{\sf SNR}_n|\bar{M}_n>0}(\gamma ;\lambda_n, \alpha, m, \zeta_n )\left(1-\exp(-\lambda L(R,\theta_n,R_{\mathcal{A}}))\right)\mbox{.}
\end{equation}
\begin{IEEEproof}
See Appendix \ref{appen:cov_prob_SNR}.
\end{IEEEproof}

The SNR-coverage probability derived in Corollary \ref{cor1} is more tractable than the SIR coverage probability in Theorem 1 because the Laplace transform of the aggregated interference power can be ignored. The SNR-coverage probability is determined by the nearest distance distribution $f_{D_{n,1}|{\bar M}_n>0}(r)$ and the fading distribution. The following example is a special case of the SNR-coverage when the fading follows the Rayleigh distribution. 

{\bf Example 2:} When $m=1$, the coverage probability in the noise-limited regime is reduced to 
\begin{align}
    \left(1-\exp(-\lambda_n L(R,\theta_n,R_{\mathcal{A}}))\right)\int_{d_{\rm min}(\theta_n)}^{d_{\rm max}}  
   \exp\left(-\frac{\sigma^2}{PG_{n,1}}\gamma r^{\alpha}\right)
   f_{D_{n,1}|\bar{M}_n>0} (r)    dr    \mbox{.}
\end{align}

\vspace{0.3cm}
To verify the analytical expression for the SNR-coverage probability, we compare it with simulation results for different values of noise power. In our simulations, we set $P$ = 40dBm, $G_{n,1}$= 30dBi, $R_h=500$, $\omega_{min}=10^{\circ}$, $\theta_n=\frac{ \pi}{2}$, $\lambda = 0.005$, $\alpha =2$, and $m=1$. As depicted in Figure \ref{fig:SNR}, our analytical expression for the SNR-coverage probability is exact for different amount of the noise power, which is computed by the multiplication of noise spectral density (-174dBm/Hz), receiver noise figure (11dB), and the bandwidth. 

Figure \ref{fig:SNR} also verifies that the SIR coverage probability derived in Theorem 1 can be a very tight approximation for the SINR coverage probability, which takes into account the noise term and the aggregated interference power. From this justification, we will keep focusing on the SIR-based coverage analysis for multi-orbit satellite networks in the sequel.




\section{  Coverage Analysis for Multi-Orbit Satellite Networks}\label{sec:multiple_orbits}

In this section, we extend our SIR-based coverage probability derived in a single-orbit satellite network to a multi-orbit satellite network to examine the synergistic gains of harnessing multi-orbits. Intuitively, exploiting a multi-orbit network can provide a macro-diversity gain by selecting the best satellite across different orbits in an opportunistic manner. We shall quantify this macro-diversity gain in terms of the SIR-coverage probability by extending the result derived in the previous section.

We consider a communication scenario in which a downlink user can opportunistically communicate by selecting the satellite that yields the maximum instantaneous SIR. In this scenario, from the definition in \eqref{eq:max_SIR}, the SIR-coverage probability can be defined as
\begin{align}\label{eq:max_SIR2}
P^{\sf cov}_{{\sf SIR}_{\rm max}}\left(\gamma; \{\lambda_n\}_{n=1}^{N}, \alpha, m, \{\zeta_n\}_{n=1}^{N}\right) = \mathbb{P}\left[\max_{n\in [N]}{\sf SIR}_n \geq \gamma\right].
\end{align}




The following theorem states the SIR-coverage when a user communicates with the nearest satellite in the orbit that provides the maximum SIR.

\theorem\label{theo:orbit_diversity} Suppose the satellite network comprises of $N$ orbits. For given orbit geometry parameters $\{\zeta_n\}_{n=1}^{N}$ and densities $\{\lambda_n\}_{n=1}^{N}$, the SIR-coverage probability for user at ${\bf u}$ is given by
\begin{align}\label{eq:orbit_div_decond}
   &P^{\sf cov}_{{\sf SIR}_{\rm max}}\left(\gamma; \{\lambda_n\}_{n=1}^{N}, \alpha, m, \{\zeta_n\}_{n=1}^{N}\right)= \nonumber\\&P^{\sf cov}_{{\sf SIR}_{\rm max}|{\bar M}_n >0}\left(\gamma; \{\lambda_n\}_{n=1}^{N}, \alpha, m, \{\zeta_n\}_{n=1}^{N}\right) \times\prod_{n=1}^N(1-\exp(-\lambda_n L(R,\theta_n,R_{\mathcal{A}}))),
\end{align}
where
\begin{align}\label{eq:orbit_div}
&P^{\sf cov}_{{\sf SIR}_{\rm max}|{\bar M}_n >0}\left(\gamma; \{\lambda_n\}_{n=1}^{N}, \alpha, m, \{\zeta_n\}_{n=1}^{N}\right) = \nonumber\\
&1-\prod_{n=1}^N\left(1-  \int_{d_{min}(\theta_n)}^{d_{\rm max}}  
   \sum_{t=0}^{m-1}\frac{(-mr^{\alpha})^t}{t!} \frac{d^t\mathcal{L}_{I_n}(s)}{ds^t}\bigg{|}_{s =m\gamma r^{\alpha}} f_{D_{n,1}|\bar{M}_n>0} (r)     {\rm d}r \right).
\end{align}

\begin{IEEEproof}
See Appendix \ref{appen:orbit_diversity}.
\end{IEEEproof}

Theorem 2 elucidates that the SIR-coverage improves regardless of the orbit geometries $\{\zeta_n\}_{n=1}^{N}$ and densities $\{\lambda_n\}_{n=1}^{N}$ as increasing the number of orbits $N$. This is because the product term in $P^{\sf cov}_{{\sf SIR}_{\rm max}|{\bar M}_n >0}\left(\gamma; \{\lambda_n\}_{n=1}^{N}, \alpha, m, \{\zeta_n\}_{n=1}^{N}\right)$ and $\prod_{n=1}^N(1-\exp(-\lambda_n L(R,\theta_n,R_{\mathcal{A}})))$ decreases as $N$ increases. As a result, we can conclude that exploiting more orbits is always beneficial to improving the downlink coverage performance. Nonetheless, the amount of the coverage enhancement is highly dependent on the orbit geometry parameters $\{\zeta_n\}_{n=1}^{N}$. One can optimize the orbit parameters to maximize the SIR-coverage probability by finding a more tractable form of the analytical expression for the SIR-coverage probability. The current form in Theorem 2 is not suitable to optimize the parameters due to complicated integrals and the derivatives of the Laplace transform. We remain this problem as future work. Instead, we show the impact of the orbit geometry via simulations.    


\begin{figure}[t]
\begin{center}
\includegraphics[scale=0.75]{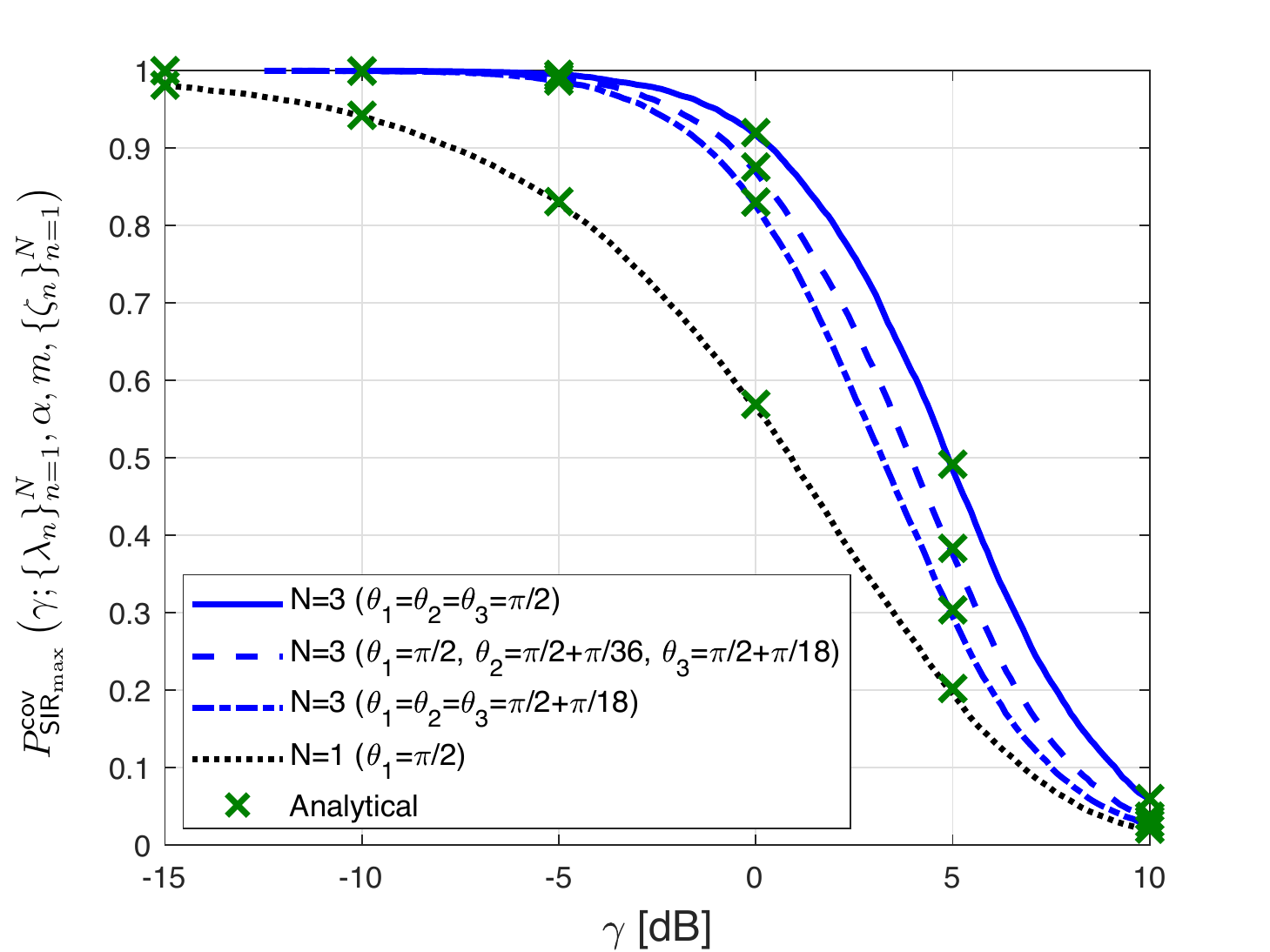}
\end{center}
\vspace{-0.5em} 
\caption{Coverage probability under orbit selection diversity. We set the parameters $R_{\sf h} = 500$, $\omega_{\rm min}=10^{\circ}$, $\lambda = 0.005$, $\alpha = 2$, $m = 1$, and $\theta_n = \frac{\pi}{2}$ for $n\in[N]$.}
\label{fig:orbit_div} 
\end{figure}

{\bf Remark 3 (Special Case):} By choosing $N=1$ in  \eqref{eq:orbit_div}, we can recover the SIR-coverage probability in \eqref{eq:cov_prob}. Hence, the coverage probability in \eqref{eq:orbit_div} is a generalization of the SIR-coverage probability for the single-orbit network.

{\bf Validation:} To validate the orbit selection diversity gain, we evaluate the SIR-coverage probability when increasing the number of orbits $N$. For simulations, we set the network parameters as $R_{\sf h} = 500$, $\omega_{\rm min}=10^{\circ}$, $\lambda = 0.005$, $\alpha = 2$, and $m = 1$. In order to investigate the effect of the number of orbits, we fix all $\theta_n = \frac{\pi}{2}$ for all $n$, while choosing the azimuth angle $\phi_n$ randomly from [0,$\pi$] since $\phi_n$ does not affect the length of the visible trajectory. This polar angle provides the maximum length of the satellite's visible trajectory to the user at ${\bf u}=(0,0,R_{\sf E})$. As can be seen in Figure \ref{fig:orbit_div}, the coverage performance keeps increasing as the number of orbits increases.

\begin{figure}[t]
\begin{center}
\includegraphics[scale=0.75]{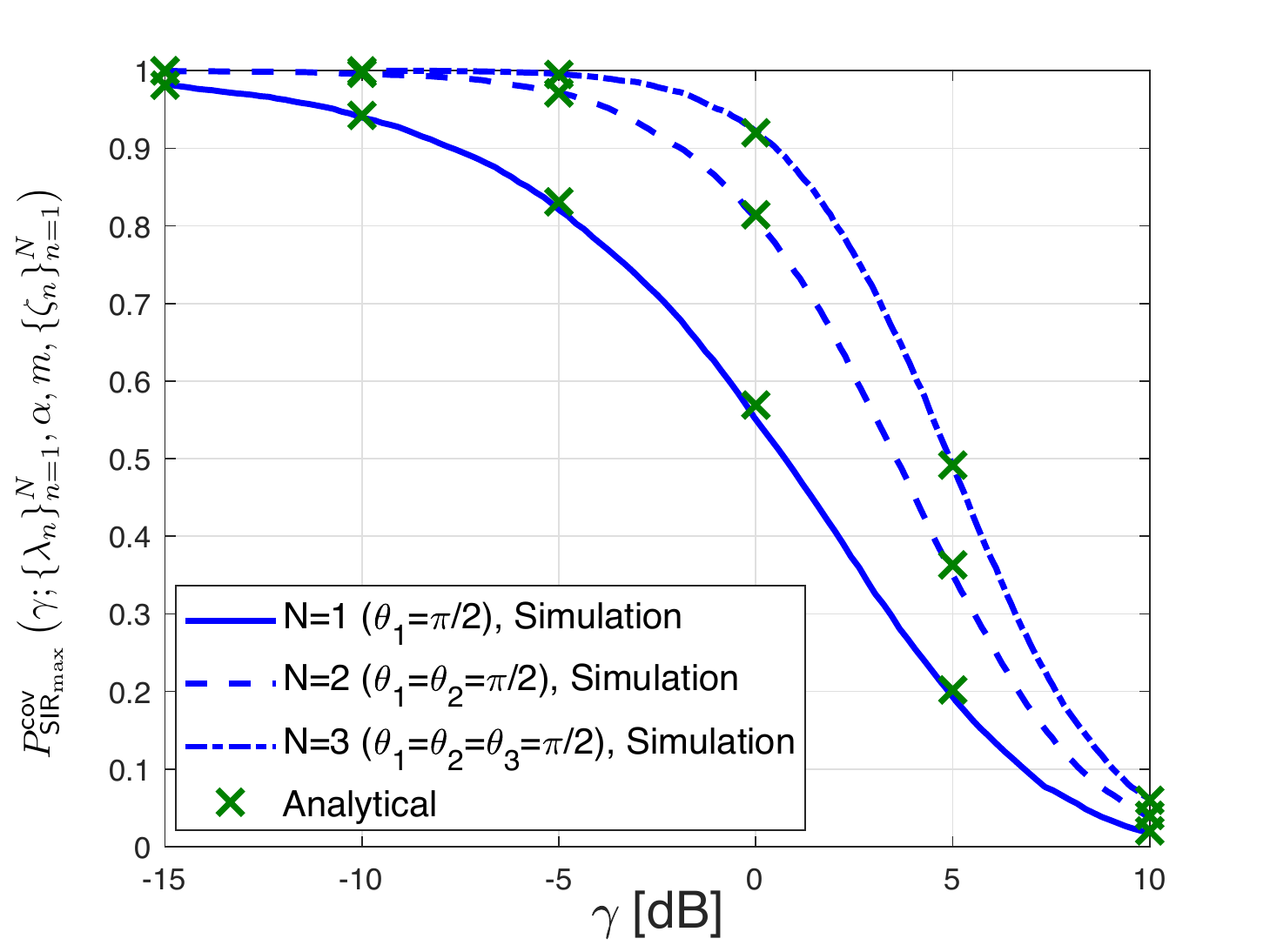}
\end{center}
\vspace{-0.5em} 
\caption{Comparison of coverage probabilities for $N=1$ and $N=3$ with different combinations of $\theta_n$ under $R_{\sf h} = 500$, $\omega_{\rm min}=10^{\circ}$, $\lambda = 0.005$, $\alpha = 2$, and $m = 1$.}
\label{fig:multiple_orbit} 
\end{figure}

We also verify the effect of the orbit geometries for different polar angles under the multi-orbit networks. When $N=3$, as can be seen in Figure \ref{fig:multiple_orbit}, the maximum coverage is achieved when the polar angles of three orbits are selected as $\theta_n = \frac{\pi}{2}$ for $n\in \{1,2,3\}$, which provides the best SIR coverage under single-orbit network as shown in Figure \ref{fig:effect_theta}. Further, by comparing the case under $N=1$ with $\theta_1=\frac{\pi}{2}$ and that under $N=3$ with $\theta_n=\frac{\pi}{2}+\frac{\pi}{18}$, we can observe that the diversity gain can overcome the degradation on coverage probability by tilted orbits. Therefore, we can conclude that exploiting more orbits for opportunistic communications benefits the diversity gain stemming from using more orbits. 


\section{Conclusion}\label{sec: Section6}
This paper analyzed the visibility and the coverage probability under a new framework for LEO satellite networks highlighting the orbit geometry. First, we derived the contact distance distribution between the user and its nearest satellite on a given orbit by adopting stochastic geometry frameworks. Then, we obtained the analytical expressions of coverage probability under the single-orbit network as a function of network parameters. We extended this result to multi-orbit networks to investigate the benefits of opportunistic diversity gain. With numerical experiments, we verified that our analytical expressions are well-matched with the simulation experiments and provide intuitions on how network parameters are related to the coverage probability. 

In this work, we provided the user's location-specific coverage performance. Potential research directions include the modeling orbits using the Poisson line process to generalize the realistic network environment. Another interesting approach to extending the coverage analysis of orbiting LEO satellites is considering heterogeneous satellite networks under multi-altitudes, or different transmit power levels. 

\appendices
\section{Proof of Lemma \ref{lem1}}\label{appen:lengthonvisiblesurface}
When $|\theta_n-\frac{\pi}{2}| > \arccos{\frac{R_{\mathcal{A}}}{R}}$, $\zeta_n$ does not pass $\mathcal{A}$, and $L(R,\theta_n,R_{\mathcal{A}})=0$. For $|\theta_n-\frac{\pi}{2}| \leq \arccos{\frac{R_{\mathcal{A}}}{R}}$, let us assume $\phi_n = 0$ without loss of generality since $L(R,\theta_n,R_{\mathcal{A}})$ is invariant over the azimuth angle, $\phi_n$. Under this assumption, $\zeta_n$ is the intersection of $x^2+y^2+z^2 = R^2$ and $\sin\theta_n x+\cos\theta_n z=0$ under the Cartesian coordinate system. We can obtain \eqref{eq:lemma1} by multiplying the radius of orbits, $R$, and the vertex angle of the isosceles triangle whose vertices are the center of the Earth and the intersecting points of $\zeta_n$ and $R_{\mathcal{A}}$, $\arccos(\eta(R,\theta_n,R_{\mathcal{A}}))$, since $\mathcal{A}$ is given as \eqref{eq:visible_surface_A}.

\section{Proof of Lemma \ref{lem:neares_satellite_1}}\label{appen:nearest_satellite_1} 
We first compute the probability that $D_{n,1}$ is larger than $r$ conditioned on $\bar{M}_n>0$. Let $\mathcal{A}_{r}$ be the subset of $\mathcal{A}$ that the distance from $\mathbf{u}$ is less than $r$. The conditional CCDF of $D_{n,1}$ conditioned on $\bar{M}_n>0$ is
\begin{align}\label{eq:appen_b_proof}
F^{c}_{D_{n,1}|M_n>0}(r)&=\mathbb{P}[D_{n,1}>r|\bar{M}_n>0] \nonumber\\&= \mathbb{P}[\Phi_{n}(\mathcal{A}_r\cap \bar{\zeta}_n)=0|\Phi_{n}(\bar{\zeta}_n)>0]\nonumber \\
&=\frac{\mathbb{P}[\Phi_{n}(\mathcal{A}_r\cap \bar{\zeta}_n)=0,\Phi_{n}(\bar{\zeta}_n)>0]}{\mathbb{P}[\Phi_{n}(\bar{\zeta}_n)>0]}\nonumber\\
&\stackrel{(a)}=\frac{\mathbb{P}[\Phi_{n}(\mathcal{A}_r\cap \bar{\zeta}_n)=0]\mathbb{P}[\Phi_{n}(\bar{\zeta}_n/(\mathcal{A}_r\cap \bar{\zeta}_n))>0]}{\mathbb{P}[\Phi_{n}(\bar{\zeta}_n)>0]}\nonumber\\
&=\frac{\mathbb{P}[\Phi_{n}(\mathcal{A}_r\cap \bar{\zeta}_n)=0](1-\mathbb{P}[\Phi_{n}(\bar{\zeta}_n/(\mathcal{A}_r\cap \bar{\zeta}_n))=0])}{1-\mathbb{P}[\Phi_{n}(\bar{\zeta}_n)=0]}\nonumber\\
& \stackrel{(b)}= \frac{\exp(-\lambda|\mathcal{A}_r\cap \bar{\zeta}_n|)(1-\exp(-\lambda|\bar{\zeta}_n/(\mathcal{A}_r\cap \bar{\zeta}_n)|))}{1-\exp(-\lambda|\bar{\zeta}_n|)}\nonumber\\
& = \frac{\exp(-\lambda|\mathcal{A}_r\cap \bar{\zeta}_n|)-\exp(-\lambda|\bar{\zeta}_n|)}{1-\exp(-\lambda|\bar{\zeta}_n|)}\nonumber\\
&\stackrel{(c)} = \frac{\exp(-\lambda|\mathcal{A}_r\cap \bar{\zeta}_n|)-\exp(-\lambda L(R,\theta_n,R_{\mathcal{A}}))}{1-\exp(-\lambda L(R,\theta_n,R_{\mathcal{A}}))}\mbox{,}
\end{align}
where (a) follows from the independence of non-overlapping PPPs, and (b) is by the void probability of PPP. Also (c) comes from Lemma \ref{lem1}. 
The length of the intersection of $\bar{\zeta}_n$ and $\mathcal{A}_r$ is invariant over $\phi_n$. In order to calculate the length of this intersecting arc, we set $\phi_n=0$ without loss of generality. By solving $\sin{\theta_n}x   + \cos{\theta_n}z = 0$, $x^2+y^2+z^2 = R^2$, and $x^2+y^2+(z-R_{\sf E})^2 < r^2$, we can obtain that
\begin{equation}\label{eq:appen_b_length}
    |\mathcal{A}_r\cap \bar{\zeta}_n| = R\arccos\left(\eta\left(R,\theta_n,\frac{R^2+R_{\sf E}^2-r^2}{2R_{\sf E}}\right)\right)\mbox{,}
\end{equation}
where $\eta(R,\theta,h)$ is defined in \eqref{eq:kappa}. By plugging \eqref{eq:appen_b_length} into \eqref{eq:appen_b_proof}, \eqref{eq:ccdf_t1} is proved.

We obtain the conditional PDF as \eqref{eq:pdf_t1} by taking derivative with respect to $r$ on \eqref{eq:ccdf_t1}.

\section{Proof of Lemma \ref{lem:Laplace_interference} }\label{appen:Laplace_interference}

By conditioning $D_{n,1} = r$, the normalized interference becomes 
\begin{align}
     \sum_{\mathbf{x}_{n,i}\in\Phi_{n}\cap\mathcal{A}_r^c}\bar{G}_{I}H_{n,i}\|\mathbf{x}_{n,i}-\mathbf{u}\|^{-\alpha}\mbox{,}\nonumber
\end{align}
where $\mathcal{A}_r^c = \mathcal{A}\setminus \mathcal{A}_r$. Then, the conditional Laplace transform of $I_n$ under $D_{n,1}=r$ is
\begin{align}
    \mathcal{L}_{ I_{n}|D_{n,1}=r}(s)
    & = \mathbb{E}\left[e^{-s I_n}|D_{n,1}=r       \right]\nonumber\\
    & = \mathbb{E}\left[\prod_{\mathbf{x}_{n,i}\in\Phi_{n}\cap\mathcal{A}_r^c}e^{-s\bar{G}_IH_{n,i}\|\mathbf{x}_{n,i}-\mathbf{u}\|^{-\alpha}} \right]\nonumber\\
    & \stackrel{(a)}= \exp\left(-\lambda\int_{v\in\mathcal{A}_r^c}\left( 1-\mathbb{E}\left[ e^{-s\bar{G}_IH_{n,i}v^{-\alpha}}    \right] \right)   dv       \right)\nonumber\\
    & \stackrel{(b)}= \exp\left(-\lambda\int_{v\in\mathcal{A}_r^c} 1 - \frac{1}{\left(1+\frac{s\bar{G}_I v^{-\alpha}}{m}\right)^m} dv                 \right)\nonumber \\
    & \stackrel{(c)}=\exp\left(-\frac{\lambda}{RR_{\sf E}^2\sin^2(\theta_n)}\times\right.\nonumber\\
    &\left.\int_{r}^{{d_{\rm max}}}\left[ 1 -\left(1+\frac{s\bar{G}_I {u}^{-\alpha}}{m}\right)^{-m}\right] \frac{2u(-u^2+R^2+R_{\sf E}^2)}{ \sqrt{1-\left(\eta\left(R,\theta_n,\frac{-u^2+R^2+R_{\sf E}^2}{2R_{\sf E}}\right)\right)^2}} du                 \right) \mbox{,}\nonumber
\end{align}
where (a) follows from the probability generating functional (PGFL) of PPP, (b) comes from deconditioning the identical independent Nakagami-$m$ fading, and (c) is obtained by change of variables with respect to the distance from the user.

\section{Proof of Theorem \ref{theo:cov_prob}}\label{appen:cov_prob}
The conditional coverage probability conditioned on $D_{n,1} = r$ and $\bar{M}_n>0$ is given by
\begin{align}
P^{\sf cov}_{ {\sf SIR}_n|{\bar M}_n >0 }(\gamma  ;\lambda_n, \alpha, m, \zeta_n )
= \mathbb{E}\left[\mathbb{P}[H_{n,1} \geq D_{n,1}^{\alpha}\gamma I_n|D_{n,1}=r ]\right]\mbox{.}\nonumber
\end{align}
Since $H_{n,1}$ is the Nakagami-$m$ random variable, the CCDF of $H_{n,1}$ is 
\begin{equation}
    \mathbb{P}[H_{n,1}\geq x] = e^{-mx}\sum_{t=0}^{m-1}\frac{(mx)^t}{t!}\mbox{.}\nonumber
\end{equation}
By leveraging this relation, we can obtain the coverage probability as
\begin{align}
&  \mathbb{E}\left[\sum_{t=0}^{m-1}\frac{m^t\gamma^tD_{n,1}^{\alpha t}}{t!}(I_n)^t e^{-m\gamma D_{n,1}^{\alpha} I_n}\bigg{|} D_{n,1}=r\right]\nonumber\\
&\stackrel{(a)}= \mathbb{E}\left[   \sum_{t=0}^{m-1}\frac{m^t r^{\alpha t}}{t!}(-1)^t \frac{d^t\mathcal{L}_{I_n|D_{n,1}=r}(s)}{ds^t}\bigg{|}_{s =m\gamma r^{\alpha}}\right]\nonumber\\
&\stackrel{(b)}= \int_{d_{\rm min}(\theta_n)}^{d_{\rm max}}  
   \sum_{t=0}^{m-1}\frac{(-m r^{\alpha })^t}{t!} \frac{d^t\mathcal{L}_{I_n|D_{n,1}=r}(s)}{ds^t}\bigg{|}_{s =m\gamma r^{\alpha}}
   f_{D_{n,1}|\bar{M}_n>0} (r)    dr \mbox{,}
\end{align}
where (a) is obtained by the derivative property of the Laplace transform and (b) comes from taking the expectation over the contact distance distribution. The coverage probability, \eqref{eq:cov_prob_prob}, is obtained by marginalizing the visibility probability.

\section{Proof of Theorem \ref{cor1}}\label{appen:cov_prob_SNR}
As in the proof of Theorem \ref{theo:cov_prob}, we compute the conditional coverage probability under $D_{n,1}=r$ and $\bar{M}_n>0$. The conditional coverage probability becomes
\begin{align}
&P^{\sf cov}_{ {\sf SNR}_n|{\bar M}_n >0 }(\gamma  ;\lambda_n, \alpha, m, \zeta_n )\nonumber\\
&= \mathbb{E}\left[\mathbb{P}\left[H_{n,1} \geq \gamma D_{n,1}^{\alpha} \frac{\sigma^2}{PG_{n,1}}\right]\bigg{|}D_{n,1}=r \right]\nonumber\\
    &\stackrel{(a)}=\mathbb{E}\left[\exp\left(-m\frac{\sigma^2}{PG_{n,1}}\gamma r^{\alpha}\right)\sum_{t=0}^{m-1}\frac{(m\gamma D_{n,1}^{\alpha}\sigma^2)^t}{t!(PG_{n,1})^t}\bigg{|}D_{n,1}=r\right]\nonumber\\
    &\stackrel{(b)}=\int_{d_{\rm min}(\theta_1)}^{d_{\rm max}}  
   \exp\left(-m\frac{\sigma^2}{PG_{n,1}}\gamma r^{\alpha}\right)\sum_{t=0}^{m-1}\frac{\left(m\gamma r^{\alpha}\sigma^2\right)^t}{t!(PG_{n,1})^t}
   f_{D_{n,1}|\bar{M}_n>0} (r)    dr \mbox{,}\nonumber
\end{align}
where (a) comes from by deconditioning $H_{n,1}$ which is the Nakagami-$m$ random variable, and (b) is obtained by taking the expectation by leveraging the nearest distance distribution in \eqref{eq:pdf_t1}.

\section{Proof of Theorem \ref{theo:orbit_diversity}}\label{appen:orbit_diversity}

Let us consider $N$ random variables, $X_1,X_2,\ldots,X_N$, which are mutually independent.  
\begin{align}\label{eq:mut_rv_lar}
    \mathbb{P}[\max(X_1,X_2,\ldots,X_N)>\gamma]&= 1- \mathbb{P}[\max(X_1,X_2,\ldots,X_N)<\gamma]\nonumber\\
    & = 1-\prod_{n=1}^N(1-\mathbb{P}[X_n>\gamma])\mbox{.}
\end{align}
\eqref{eq:mut_rv_lar} shows that the CCDF of $\max(X_1,X_2,\ldots,X_N)$ is $1-(\mbox{Products of the CDF of $X_n$})$.

Since the sum of the received power from each $\zeta_n$ are mutually independent, the coverage probability under the orbit selection diversity scheme becomes
\begin{align}\label{eq:orbit_div_pro}
&P^{\sf cov}_{{\sf SIR}_{\rm max}|{\bar M}_n >0}\left(\gamma; \{\lambda_n\}_{n=1}^{N}, \alpha, m, \{\zeta_n\}_{n=1}^{N}\right)\nonumber\\
&=1-\prod_{n=1}^{N}\left(1-P^{\sf cov}_{ {\sf SIR}_n|{\bar M}_n >0 }(\gamma  ;\lambda_n, \alpha, m, \zeta_n )\right)\nonumber\\
&=1-\prod_{n=1}^N\left(1-  \int_{d_{min}(\theta_n)}^{d_{\rm max}}  
   \sum_{t=0}^{m-1}\frac{(-mr^{\alpha})^t}{t!} \frac{d^t\mathcal{L}_{I_n}(s)}{ds^t}\bigg{|}_{s =m\gamma r^{\alpha}} f_{D_{n,1}|\bar{M}_n>0} (r)     dr \right).
\end{align}
Then, the coverage probability, \eqref{eq:orbit_div_decond}, is obtained by deconditioning the visibility probability.

\bibliographystyle{ieeetran}
\bibliography{referenceBibs}

\begin{thebibliography}{10}
\providecommand{\url}[1]{#1}
\csname url@samestyle\endcsname
\providecommand{\newblock}{\relax}
\providecommand{\bibinfo}[2]{#2}
\providecommand{\BIBentrySTDinterwordspacing}{\spaceskip=0pt\relax}
\providecommand{\BIBentryALTinterwordstretchfactor}{4}
\providecommand{\BIBentryALTinterwordspacing}{\spaceskip=\fontdimen2\font plus
\BIBentryALTinterwordstretchfactor\fontdimen3\font minus
  \fontdimen4\font\relax}
\providecommand{\BIBforeignlanguage}[2]{{%
\expandafter\ifx\csname l@#1\endcsname\relax
\typeout{** WARNING: IEEEtran.bst: No hyphenation pattern has been}%
\typeout{** loaded for the language `#1'. Using the pattern for}%
\typeout{** the default language instead.}%
\else
\language=\csname l@#1\endcsname
\fi
#2}}
\providecommand{\BIBdecl}{\relax}
\BIBdecl

\bibitem{chen2020vision}
S.~Chen, Y.-C. Liang, S.~Sun, S.~Kang, W.~Cheng, and M.~Peng, ``Vision,
  requirements, and technology trend of 6g: How to tackle the challenges of
  system coverage, capacity, user data-rate and movement speed,'' \emph{IEEE
  Wireless Communications}, vol.~27, no.~2, pp. 218--228, 2020.

\bibitem{zhang20196g}
Z.~Zhang, Y.~Xiao, Z.~Ma, M.~Xiao, Z.~Ding, X.~Lei, G.~K. Karagiannidis, and
  P.~Fan, ``6g wireless networks: Vision, requirements, architecture, and key
  technologies,'' \emph{IEEE Vehicular Technology Magazine}, vol.~14, no.~3,
  pp. 28--41, 2019.

\bibitem{giordani2020toward}
M.~Giordani, M.~Polese, M.~Mezzavilla, S.~Rangan, and M.~Zorzi, ``Toward 6g
  networks: Use cases and technologies,'' \emph{IEEE Communications Magazine},
  vol.~58, no.~3, pp. 55--61, 2020.

\bibitem{international2019measuring}
I.~T. Union, ``Measuring digital development: Facts and figures 2019,'' 2019.

\bibitem{baccelli2009stochastic}
F.~Baccelli and B.~Blaszczyszyn, \emph{Stochastic Geometry and Wireless
  Networks: Volume 1: THEORY}.\hskip 1em plus 0.5em minus 0.4em\relax Now
  Publishers Inc, 2009, vol.~1.

\bibitem{baccelli2006aloha}
F.~Baccelli, B.~Blaszczyszyn, and P.~Muhlethaler, ``An aloha protocol for
  multihop mobile wireless networks,'' \emph{Information Theory, IEEE
  Transactions on}, vol.~52, no.~2, pp. 421--436, 2006.

\bibitem{baccelli2009stochasticopp}
------, ``Stochastic analysis of spatial and opportunistic aloha,'' \emph{IEEE
  journal on selected areas in communications}, vol.~27, no.~7, pp. 1105--1119,
  2009.

\bibitem{NLee3}
N.~Lee, F.~Baccelli, and R.~W. Heath, ``Spectral efficiency scaling laws in
  dense random wireless networks with multiple receive antennas,'' \emph{IEEE
  Transactions on Information Theory}, vol.~62, no.~3, pp. 1344--1359, 2016.

\bibitem{andrews2011tractable}
J.~G. Andrews, F.~Baccelli, and R.~K. Ganti, ``A tractable approach to coverage
  and rate in cellular networks,'' \emph{IEEE Transactions on communications},
  vol.~59, no.~11, pp. 3122--3134, 2011.

\bibitem{dhillon2012modeling}
H.~S. Dhillon, R.~K. Ganti, F.~Baccelli, and J.~G. Andrews, ``Modeling and
  analysis of k-tier downlink heterogeneous cellular networks,'' \emph{IEEE
  Journal on Selected Areas in Communications}, vol.~30, no.~3, pp. 550--560,
  2012.

\bibitem{di2013average}
M.~Di~Renzo, A.~Guidotti, and G.~E. Corazza, ``Average rate of downlink
  heterogeneous cellular networks over generalized fading channels: A
  stochastic geometry approach,'' \emph{IEEE Transactions on Communications},
  vol.~61, no.~7, pp. 3050--3071, 2013.

\bibitem{NLee1}
N.~Lee, X.~Lin, J.~G. Andrews, and R.~W. Heath, ``Power control for d2d
  underlaid cellular networks: Modeling, algorithms, and analysis,'' \emph{IEEE
  Journal on Selected Areas in Communications}, vol.~33, no.~1, pp. 1--13,
  2015.

\bibitem{NLee2}
N.~Lee, D.~Morales-Jimenez, A.~Lozano, and R.~W. Heath, ``Spectral efficiency
  of dynamic coordinated beamforming: A stochastic geometry approach,''
  \emph{IEEE Transactions on Wireless Communications}, vol.~14, no.~1, pp.
  230--241, 2015.

\bibitem{bai2014coverage}
T.~Bai, A.~Alkhateeb, and R.~W. Heath, ``Coverage and capacity of
  millimeter-wave cellular networks,'' \emph{IEEE Communications Magazine},
  vol.~52, no.~9, pp. 70--77, 2014.

\bibitem{di2015stochastic}
M.~Di~Renzo, ``Stochastic geometry modeling and analysis of multi-tier
  millimeter wave cellular networks,'' \emph{IEEE Transactions on Wireless
  Communications}, vol.~14, no.~9, pp. 5038--5057, 2015.

\bibitem{tong2016stochastic}
Z.~Tong, H.~Lu, M.~Haenggi, and C.~Poellabauer, ``A stochastic geometry
  approach to the modeling of dsrc for vehicular safety communication,''
  \emph{IEEE Transactions on Intelligent Transportation Systems}, vol.~17,
  no.~5, pp. 1448--1458, 2016.

\bibitem{yi2019modeling}
W.~Yi, Y.~Liu, Y.~Deng, A.~Nallanathan, and R.~W. Heath, ``Modeling and
  analysis of mmwave v2x networks with vehicular platoon systems,'' \emph{IEEE
  Journal on Selected Areas in Communications}, vol.~37, no.~12, pp.
  2851--2866, 2019.

\bibitem{chetlur2017downlink}
V.~V. Chetlur and H.~S. Dhillon, ``Downlink coverage analysis for a finite 3-d
  wireless network of unmanned aerial vehicles,'' \emph{IEEE Transactions on
  Communications}, vol.~65, no.~10, pp. 4543--4558, 2017.

\bibitem{banagar2020performance}
M.~Banagar and H.~S. Dhillon, ``Performance characterization of canonical
  mobility models in drone cellular networks,'' \emph{IEEE Transactions on
  Wireless Communications}, vol.~19, no.~7, pp. 4994--5009, 2020.

\bibitem{okati2021modeling}
N.~Okati and T.~Riihonen, ``Modeling and analysis of leo mega-constellations as
  nonhomogeneous poisson point processes,'' in \emph{2021 IEEE 93rd Vehicular
  Technology Conference (VTC2021-Spring)}.\hskip 1em plus 0.5em minus
  0.4em\relax IEEE, 2021, pp. 1--5.

\bibitem{okati2022coverage}
------, ``Coverage and rate analysis of mega-constellations under generalized
  serving satellite selection,'' in \emph{2022 IEEE Wireless Communications and
  Networking Conference (WCNC)}.\hskip 1em plus 0.5em minus 0.4em\relax IEEE,
  2022, pp. 2214--2219.

\bibitem{okati2022nonhomogeneous}
------, ``Nonhomogeneous stochastic geometry analysis of massive leo
  communication constellations,'' \emph{IEEE Transactions on Communications},
  vol.~70, no.~3, pp. 1848--1860, 2022.

\bibitem{al2022next}
B.~Al~Homssi, A.~Al-Hourani, K.~Wang, P.~Conder, S.~Kandeepan, J.~Choi,
  B.~Allen, and B.~Moores, ``Next generation mega satellite networks for access
  equality: Opportunities, challenges, and performance,'' \emph{IEEE
  Communications Magazine}, vol.~60, no.~4, pp. 18--24, 2022.

\bibitem{talgat2020nearest}
A.~Talgat, M.~A. Kishk, and M.-S. Alouini, ``Nearest neighbor and contact
  distance distribution for binomial point process on spherical surfaces,''
  \emph{IEEE Communications Letters}, vol.~24, no.~12, pp. 2659--2663, 2020.

\bibitem{talgat2020stochastic}
------, ``Stochastic geometry-based analysis of leo satellite communication
  systems,'' \emph{IEEE Communications Letters}, vol.~25, no.~8, pp.
  2458--2462, 2020.

\bibitem{al2021analytic}
A.~Al-Hourani, ``An analytic approach for modeling the coverage performance of
  dense satellite networks,'' \emph{IEEE Wireless Communications Letters},
  vol.~10, no.~4, pp. 897--901, 2021.

\bibitem{al2021optimal}
------, ``Optimal satellite constellation altitude for maximal coverage,''
  \emph{IEEE Wireless Communications Letters}, vol.~10, no.~7, pp. 1444--1448,
  2021.

\bibitem{park2021coverage}
J.~Park, J.~Choi, and N.~Lee, ``A tractable approach to coverage analysis in
  downlink satellite networks,'' \emph{arXiv preprint arXiv:2111.12851}, 2021.

\bibitem{walker1970circular}
J.~G. Walker, ``Circular orbit patterns providing continuous whole earth
  coverage,'' ROYAL AIRCRAFT ESTABLISHMENT FARNBOROUGH (UNITED KINGDOM), Tech.
  Rep., 1970.

\bibitem{ganz1994performance}
A.~Ganz, Y.~Gong, and B.~Li, ``Performance study of low earth-orbit satellite
  systems,'' \emph{IEEE Transactions on Communications}, vol.~42, no. 234, pp.
  1866--1871, 1994.

\bibitem{vatalaro1995analysis}
F.~Vatalaro, G.~E. Corazza, C.~Caini, and C.~Ferrarelli, ``Analysis of leo,
  meo, and geo global mobile satellite systems in the presence of interference
  and fading,'' \emph{IEEE Journal on selected areas in communications},
  vol.~13, no.~2, pp. 291--300, 1995.

\bibitem{okati2020downlink}
N.~Okati, T.~Riihonen, D.~Korpi, I.~Angervuori, and R.~Wichman, ``Downlink
  coverage and rate analysis of low earth orbit satellite constellations using
  stochastic geometry,'' \emph{IEEE Transactions on Communications}, vol.~68,
  no.~8, pp. 5120--5134, 2020.

\bibitem{giunta2018estimation}
G.~Giunta, C.~Hao, and D.~Orlando, ``Estimation of rician k-factor in the
  presence of nakagami-$ m $ shadowing for the los component,'' \emph{IEEE
  Wireless Communications Letters}, vol.~7, no.~4, pp. 550--553, 2018.

\bibitem{koretz2009dolph}
A.~Koretz and B.~Rafaely, ``Dolph--chebyshev beampattern design for spherical
  arrays,'' \emph{IEEE transactions on Signal processing}, vol.~57, no.~6, pp.
  2417--2420, 2009.

\end{thebibliography}

\end{document}